\pdfoutput=1
\documentclass[11pt,twoside,a4paper,cmspaper,final,collab]{cms-tdr}
\def\svnVersion{e97b670-D}\def\svnDate{2021/05/19}\def\cmsCernNoTag{CERN-EP-2020-223}\def\cmsCernDate{\today}\def\cmsMessage{"Published in the Journal of High Energy Physics as \href{http://dx.doi.org/10.1007/JHEP05(2021)205}{\doi{10.1007/JHEP05(2021)205}.}"}
\begin{document}\cmsNoteHeader{SMP-18-003}

\newcommand{\tw}{\ensuremath{\PQt\PW}\xspace}
\newcommand{\Zmm}{\ensuremath{\PZ\to\PGmp\PGmm}\xspace}
\newcommand{\Zee}{\ensuremath{\PZ\to \Pep\Pem}\xspace}
\newcommand{\Zll}{\ensuremath{\PZ\to\ell\ell}\xspace}
\newcommand{\Zllvv}{\ensuremath{\PZ\to\ell\ell+\PGn\PGn}\xspace}
\newcommand{\Zvv}{\ensuremath{\PZ\to\PGn\PGn}\xspace}
\newcommand{\Zvvbar}{\ensuremath{\PZ\to\PGn\bar{\PGn}}\xspace}
\newcommand{\Wlv}{\ensuremath{\PW\to \ell\PGn}\xspace}
\newcommand{\Wmn}{\ensuremath{\PW\to \PGm\PGn}\xspace}
\newcommand{\Wen}{\ensuremath{\PW\to \Pe\PGn}\xspace}
\newcommand{\Zmmjets}{\ensuremath{\Zmm+\text{jets}}\xspace}
\newcommand{\Zlljets}{\ensuremath{\Zll+\text{jets}}\xspace}
\newcommand{\Zjets}{\ensuremath{\PZ+\text{jets}}\xspace}
\newcommand{\Wjets}{\ensuremath{\PW+\text{jets}}\xspace}
\newcommand{\Wlvjets}{\ensuremath{\PW(\ell\PGn)+\text{jets}}\xspace}
\newcommand{\Wmvjets}{\ensuremath{\PW(\PGm\PGn)+\text{jets}}\xspace}
\newcommand{\Wevjets}{\ensuremath{\PW(\Pe\PGn)+\text{jets}}\xspace}
\newcommand{\phojets}{\ensuremath{\gamma+\text{jets}}\xspace}
\newcommand{\mettrig}{\ensuremath{p_{\text{T, trig}\xspace}^{\text{miss}}\xspace}\xspace}
\newcommand{\mhttrig}{\ensuremath{H_{\text{T, trig}\xspace}^{\text{miss}}\xspace}\xspace}
\newcommand{\ptvecjet}{\ensuremath{\ptvec^{\kern1pt\text{jet}}}\xspace}
\newcommand{\zpt}{\ensuremath{\pt^{\PZ}}\xspace}
\newcommand{\mll}{\ensuremath{m_{\ell\ell}}}
\newlength\cmsTabSkip\setlength{\cmsTabSkip}{1ex}
\ifthenelse{\boolean{cms@external}}{\providecommand{\cmsTable}[1]{#1}}{\providecommand{\cmsTable}[1]{\resizebox{\textwidth}{!}{#1}}}

\cmsNoteHeader{SMP-18-003}

\title{Measurement of the \texorpdfstring{$\PZ$}{Z} boson differential production cross section using its invisible decay mode (\texorpdfstring{$\Zvvbar$}{Znn}) in proton-proton collisions at \texorpdfstring{$\sqrt{s}=13\TeV$}{sqrt(s) = 13 TeV}}

\date{\today}

\abstract{
Measurements of the total and differential fiducial cross sections for the \PZ  boson decaying into two neutrinos are presented at the LHC in proton-proton collisions at a center-of-mass energy of $13\TeV$. The data were collected by the CMS detector in 2016 and correspond to an integrated luminosity of 35.9\fbinv. In these measurements, events are selected containing an imbalance in transverse momentum and one or more energetic jets. The fiducial differential cross section is measured as a function of the \PZ boson transverse momentum. The results are combined with a previous measurement of charged-lepton decays of the \PZ boson. The measured total fiducial cross section for events with \PZ boson transverse momentum greater than 200\GeV is $3000^{+180}_{-170}$\unit{fb}.
}

\hypersetup{%
  pdfauthor={CMS Collaboration},%
  pdftitle={Measurement of the Z boson differential production cross section in the invisible decay mode in proton-proton collisions at sqrts = 13 TeV},%
  pdfsubject={CMS},%
  pdfkeywords={CMS, Z boson, cross section, invisible decay}
}

\maketitle

\section{Introduction}
\label{sec:introduction}

The precision measurements of the production of neutrino pairs via the \PZ boson is an important aspect of the Large Hadron Collider (LHC) program for testing the standard model (SM) of particle physics. They provide a reference for various other measurements in the high energy regime, especially when searching for contributions beyond the SM, where the \PZ boson production constitutes an important background process. 
Expectations for the \PZ boson cross section have been calculated up to next-to-next-to-leading 
order (NNLO) in perturbative quantum chromodynamics (QCD) and up to next-to-leading order (NLO) in 
electroweak (EW) production, which correspond to a full NLO EW theory~\cite{Denner:2009gj,Denner:2011vu,Denner:2012ts,Kallweit:2015dum} supplemented by two-loop Sudakov EW logarithms~\cite{Kuhn:2004em,Kuhn:2005gv,Kuhn:2005az,Kuhn:2007cv}. The calculations are limited by the uncertainties in parton distribution functions (PDFs), and higher-order QCD and EW corrections~\cite{Lindert:2017olm}.

Measurements of the differential \PZ boson production cross sections have been reported by both ATLAS~\cite{ATLAS_ZpT7TeV,ATLAS_ZptEta7TeV,Aad:2015auj,Aad:2019wmn} and CMS~\cite{CMS_ZpT7TeV,CMS_ZpT8TeV,CMS:2014jea,CMS_Z_13TeV} at the CERN LHC using charged leptons (electrons or muons) in the final states. In addition, ATLAS has presented differential measurements in pp collisions at $\sqrt{s}=13\TeV$ of the ratio of $\PZ \to \nu\nu$/$\PZ \to \ell\ell$~\cite{Aaboud:2017buf} and $\PZ\PZ \to \nu\nu \ell\ell$ production cross sections~\cite{Aaboud:2019lgy}. In this paper, we present the differential production cross section measurement of $\cPZ$ bosons identified via their decays to pairs of neutrinos. The \PZ boson branching fraction to neutrinos is about a factor of six times that to electrons or muons, which leads to a smaller statistical uncertainty. This can only be fully exploited at large transverse momenta of the \PZ boson ($\zpt$), above $\approx$500\GeV, where the measurement of the missing transverse momentum (\ptmiss) is sufficiently accurate. This measurement therefore complements those using the charged lepton final states and improves their precision at a higher energy scale. A significant deviation, in particular at large $\zpt$, could reveal signs of physics beyond the SM~\cite{Sirunyan:2017jix,Sirunyan:2018owy,Sirunyan:2019ctn}.

This paper presents the first inclusive, differential, and normalized fiducial cross section measurements as functions of $\zpt$, where the \PZ boson is identified via its decay to a pair of neutrinos. The neutrinos are not detected by the CMS detector, but are reconstructed indirectly through the transverse momentum imbalance in the event. We use events 
with energetic jets and large $\ptmiss$, where the jets mainly arise from the fragmentation and hadronization of quarks or gluons that are produced in the hard scattering process as initial-state radiation. The analysis is based on a data sample of proton-proton ($\Pp\Pp$) collisions at a center-of-mass energy of $13\TeV$, corresponding to an integrated luminosity of $35.9\pm0.9$\fbinv collected with the CMS detector at the LHC in 2016.

This paper is organized as follows. A brief overview of the CMS detector is given in Section~\ref{sec:detector}. Information about the definition of objects used in the analysis and the event selection is summarized in Section~\ref{sec:selection}. Event simulations with various Monte Carlo (MC) generators are discussed in Section~\ref{sec:simulation}. Section~\ref{sec:signal-extraction} explains the signal-extraction strategy and Section~\ref{sec:cross_section} discusses the total, differential, and normalized fiducial cross section measurements. A combined analysis of the current measurements with those from charged leptons is presented in Section~\ref{sec:combination}. Finally, we summarize our results in Section~\ref{sec:summary}. 
Tabulated results are available in the HepData database~\cite{hepdata}.

\section{The CMS detector and event reconstruction}
\label{sec:detector}

The central feature of the CMS apparatus is a superconducting solenoid of 6\unit{m} internal diameter, providing a magnetic field of 3.8\unit{T}. A silicon pixel and strip tracker, a lead tungstate crystal electromagnetic calorimeter (ECAL), and a brass and scintillator hadron calorimeter (HCAL), each composed of a barrel and two end sections reside within the solenoid volume. Forward calorimeters extend the pseudorapidity ($\eta$) coverage provided by the barrel and end-section detectors. Muons are detected in gas-ionization chambers embedded in the steel flux-return yoke outside the solenoid. 
A more detailed description of the CMS detector, together with a definition of the coordinate system and kinematic variables, can be found in Ref.~\cite{Chatrchyan:2008aa}. Events of interest are selected using a two-tiered trigger system~\cite{Khachatryan:2016bia}. The first level, composed of custom hardware processors, uses information from the calorimeters and muon detectors at an output rate of $\approx$100\unit{kHz} within a fixed latency of about 4\unit{\mus}. The second level, known as the high-level trigger, consists of a farm of processors running a version of the full event reconstruction software optimized for fast processing, that reduces the event rate to $\approx$1\unit{kHz}
before data storage. Additional $\Pp\Pp$ interactions to the studied collision that take place in the same or nearby bunch crossings are referred to as pileup.

Events are reconstructed using a particle-flow (PF) algorithm~\cite{Sirunyan:2017ulk}, which combines information from the tracker, calorimeters, and muon systems to reconstruct and identify charged and neutral hadrons, photons, muons, and electrons. The definitions in the following analysis are based on the reconstructed PF candidates, and incorporate additional algorithms and requirements to optimize the selected objects.

Jets are reconstructed by clustering PF candidates using the anti-\kt algorithm~\cite{antikt,Cacciari:2011ma} with a distance parameter $R=0.4$. The jet energies are calibrated in the simulation, and separately in data, accounting for energy deposits of neutral particles from pileup and any nonlinear detector response~\cite{Khachatryan:2016kdb,CMS-DP-2020-019}. 
A charged-hadron subtraction technique, which removes the energy of charged hadrons not originating from the event primary vertex (PV)~\cite{Sirunyan:2020foa}, is applied to mitigate the effect of pileup. The PV is defined as the vertex with the largest value of summed physics-object $\pt^2$. Here, the physics objects are the jets clustered using the jet finding algorithm~\cite{antikt,Cacciari:2011ma} with the tracks assigned to the vertex as inputs, and the associated $\ptmiss$, which is the negative vector $\pt$ sum of those jets. The estimation of $\ptmiss$ is improved by propagating the energy correction to the jet four vectors into the sum, as described in Ref.~\cite{Sirunyan:2019kia}. 

Jets originating from \PQb quarks are identified (\PQb{}-tagged) using a multivariate algorithm, referred to as the combined secondary vertex algorithm (CSVv2)~\cite{Sirunyan:2017ezt}. The \PQb tagging working point used in this analysis has a tagging efficiency of $\approx$65\% as measured for jets originating from the hadronization of \PQb quarks in top quark pair (\ttbar) events, with a corresponding mistag rate for jets that originate from the hadronization of light flavor quarks of $\approx$1\%~\cite{Sirunyan:2017ezt}.

Electron candidates are reconstructed by matching clusters of energy in the ECAL to tracks in the silicon tracker~\cite{Khachatryan:2015hwa}. Clusters compatible with electromagnetic deposition become seeds for electron tracks by back-propagating their trajectories from the calorimeter to the tracker. Tracker seeds are also created from existing tracks by extrapolating their trajectories to the ECAL surface and associating them to PF clusters. After seeds are created, track reconstruction is performed using a dedicated fitting algorithm that includes bremsstrahlung photons that are compatible with originating from an electron track. Additional requirements are applied to reject electrons created in photon conversions in tracker material or jets misreconstructed as electrons. Electron identification criteria rely on observables sensitive to bremsstrahlung along the electron trajectory, the geometrical and momentum-energy matching between the electron trajectory and the associated energy deposit in ECAL, as well as the distribution of energy in the shower and its association with the PV. 

Muon candidates are reconstructed in the central tracking system alone or by combining charged tracks in the muon detector with trajectories in the central tracker~\cite{Sirunyan:2018fpa}. Identification criteria based on the number of measurements in the tracker and in the muon system, the fit quality of the muon track, and its consistency with its origin from the PV are imposed on the muon candidates to reduce the misidentification rate. 

Prompt charged leptons are usually isolated, whereas misidentified leptons are often accompanied by charged hadrons or neutral particles. Leptons also arise from a secondary vertex if they are decay products of bottom or charm hadrons. To identify prompt charged leptons we require electrons and muons to satisfy an isolation criterion. An isolation variable is defined by the \pt sum over charged PF candidates associated to the PV and neutral PF particles within a cone around the lepton of radius $\Delta R = \sqrt{\smash[b]{(\Delta\eta)^2+(\Delta\phi)^2}} = 0.4$, excluding the lepton. Here $\Delta\phi$ and $\Delta\eta$ refer to the differences in the $\phi$ (azimuth) and $\eta$ 
variables of the PF candidate to the charged-lepton candidate. Isolation cannot be larger than a given maximum value~\cite{Sirunyan:2017jix}. To mitigate the effect of pileup on this variable, a correction is implemented based on the total event occupancy~\cite{Cacciari:2007fd}.

Electron and muon candidates must pass certain identification criteria to be selected in the analysis. For the ``loose'' identification and isolation, they must satisfy $\pt > 10\GeV$ and $\abs{\eta} < 2.5$ (2.4) for electrons (muons). At the final stage of the lepton selection, the ``tight'' working points, following the definitions provided in Ref.~\cite{Khachatryan:2015hwa,Sirunyan:2018fpa}, are chosen for the identification criteria, including requirements on the impact parameter of the candidates with respect to the PV and their isolation with respect to other particles in the event~\cite{Sirunyan:2018egh}.

The $\tau$ leptons that decay to hadrons (\tauh) are identified using the ``hadron-plus-strips'' algorithm~\cite{Khachatryan:2015dfa}. This algorithm constructs candidates seeded by PF jets that are consistent with \Pgt lepton decay with one or three charged pions. In the single charged pion decay mode, the presence of neutral pions is detected by reconstructing their photonic decays. Mistagged jets originating from non-\Pgt decays are rejected by a discriminator that takes into account the pileup contribution to the neutral component of the \tauh decay~\cite{Sirunyan:2018pgf}. In addition, decay candidates are, in similarity to electron and muon candidates, required to satisfy an isolation criterion as described in Ref.~\cite{Khachatryan:2015dfa}.

Photon candidates are reconstructed~\cite{CMS:EGM-14-001} from clusters in the ECAL that are required to be isolated. 
The energy deposition in the HCAL tower closest to the seed of the ECAL supercluster~\cite{Khachatryan:2015hwa} assigned 
to the photon is required to contain $<$5\% of the energy deposited in the ECAL. The photon isolation is defined by the sum over scalar $\pt$ in a cone centered around the photon momentum vector with a radius of $\Delta R = 0.3$, 
that excludes the contribution of the photon candidate itself. Corrections for pileup effects are applied to the isolation criteria and depend on the $\eta$ of the photon.

\section{Event selection}
\label{sec:selection}

We define a signal region (SR) and two control regions (CR). The CRs make use of single-muon and single-lepton events. While the number of signal events is extracted from the SR, the CRs are used to constrain the dominant \Wlvjets background. 

\subsection{Signal region}

For the signal region events are selected using a dedicated triggers designed to select events with large $\mettrig$ and $\mhttrig$ as calculated with the PF algorithm in the trigger. The observable $\mettrig$ corresponds to the magnitude of the vector $\ptvec$ sum of all PF candidates reconstructed at the trigger level, while the $\mhttrig$ is the scalar sum of jet $\pt$s with $\pt>20\GeV$ and $\abs{\eta}<5.0$ reconstructed at the trigger level. The energy fraction attributed to neutral hadrons in these jets is required to be smaller than 0.9 to suppress events with jets originating from detector noise. The trigger efficiency is measured to be 97\% for events passing the analysis selection with $\ptmiss>250\GeV$, and becomes 100\% for events with $\ptmiss>350\GeV$~\cite{Sirunyan:2017jix} with respect to the trigger thresholds of 110 or 120\GeV on $\mettrig$ and $\mhttrig$ that depend on the data-taking period.

Events are required to have $\ptmiss>250\GeV$, and the leading jet in the event is required to have $\pt>100\GeV$ and $\abs{\eta}<2.5$. The leading jet is also required to have at least 10\% of its energy associated with charged particles and less than 80\% to neutral hadrons. This selection helps to reduce beam-induced background events. In addition, the analysis employs various event filters to reduce large misreconstructed $\ptmiss$ backgrounds not originating from beam-beam collisions~\cite{Sirunyan:2019kia}. The background from $\Wlvjets$ is suppressed by imposing a veto on events containing one or more loosely identified electron or muon with $\pt>10\GeV$, or $\tau$ leptons with $\pt>18\GeV$. Events that contain an isolated photon with $\pt>15\GeV$ and $\abs{\eta}<2.5$ are also vetoed to reduce the \phojets\ background. To reduce the contamination from top quark backgrounds, events are rejected if they contain an identified \PQb quark jet candidate with $\pt>20\GeV$ and $\abs{\eta}<2.4$. Finally, production of multijet background events from QCD processes with $\ptmiss$ arising from mismeasurements of the jet momenta is suppressed by requiring the minimum 
azimuthal angle between the $\ptvecmiss$ direction and each of the first four jets with \pt greater than $30$\GeV ($\Delta\phi(\ptvecjet,\ptvecmiss)$) to be larger than 0.5 radians. The selection requirements in the SR are summarized in Table~\ref{tab:selection}.

\begin{table}[htbp]
\centering
\topcaption{\label{tab:selection}
  Summary of the signal region definition.}
\cmsTable{
  \begin{tabular}{lcc}
    \hline
    Variable                               & Selection                              & To suppress background from \\
    \hline
    Electron veto                          & $\pt>10\GeV,\abs{\eta}<2.5$	           & \Zlljets,~\Wlvjets \\
    Muon veto                              & $\pt>10\GeV,\abs{\eta}<2.4$	           & \Zlljets,~\Wlvjets \\
    $\tau$ veto                            & $\pt>18\GeV,\abs{\eta}<2.3$                   & \Zlljets,~\Wlvjets \\
    Photon veto                            & $\pt>15\GeV,\abs{\eta}<2.5$                   & \phojets \\ 
    \PQb jet veto                         & CSVv2 $<$0.8484, $\pt>20\GeV,~\abs{\eta}<2.4$ & Top quark \\
    $\ptmiss$                              & $>$250\GeV                                    & QCD multijet, top quark, \Zlljets \\
    $\Delta\phi(\ptvecjet,\ptvecmiss)$     & $>$0.5 radians                                & QCD multijet \\ 
    Leading jet                            & $\pt>100\GeV$, $\abs{\eta}<2.4$               & All \\
    \hline
  \end{tabular}
}
\end{table}

\subsection{Single-muon control region}

The dedicated trigger algorithm discussed for the signal region also provides events for the single-muon CR as the identified PF muons are not considered in the computations of $\mettrig$ and $\mhttrig$. The single-muon CR is enriched in $\Wmvjets$ events and is defined using the SR criteria with two modifications. First, the muon veto is not applied, and exactly one tightly identified and isolated muon with $\pt>20\GeV$ is required~\cite{Sirunyan:2018fpa}. No additional loosely identified electrons or muons with $\pt>10\GeV$ are accepted. Second, the transverse mass of the muon-$\ptvecmiss$ system, given by $\mT = \sqrt{\smash[b]{2\ptmiss~ \pt^{\PGm} (1 - \mathrm{cos}\Delta\phi)}}$, where $\pt^{\PGm}$ is the \pt of the muon, and $\Delta\phi$ is the azimuthal angle between $\ptvec^{\PGm}$ and $\ptvecmiss$, is required to be less than 160\GeV. This suppresses background from QCD multijet events. 

\subsection{Single-electron control region}

The single-electron CR is selected from a different data set based on isolated and non isolated single-electron triggers. The trigger efficiency for the single-electron CR is measured to be $\approx$90\% for events passing the selection for electrons with $\pt>40\GeV$, and becomes fully efficient for events with electron at $\pt>100\GeV$. The single-electron CR is enriched in $\Wevjets$ events and is defined using the trigger selection described above. Events in the single-electron CR are required to contain exactly one tightly identified and isolated electron with $\pt>40\GeV$~\cite{Khachatryan:2015hwa}, and no additional loosely identified electrons or muons with $\pt>10\GeV$. Finally, the contamination from QCD multijet events is suppressed by requiring $\ptmiss>50\GeV$ and having the $\mT$ of the electron-$\ptvecmiss$ system satisfy $\mT<160\GeV$.

\section{Signal and background simulation}
\label{sec:simulation}

Multiple MC event generators are used to simulate the signal and background processes in this analysis. The simulated events are used to optimize selections, evaluate selection efficiencies and systematic uncertainties, and compute expected yields.

Simulated MC samples are produced for the $\Zjets$ and $\Wjets$ processes at next-to-leading order (NLO) in QCD using the 
\MGvATNLO~2.2.2~\cite{Alwall:2014hca} generator with up to two additional partons in the matrix element calculations. These events are corrected by vector boson $\pt$-dependent higher-order EW terms using the multiplicative prescription given in Refs.~\cite{Denner:2009gj,Denner:2011vu,Denner:2012ts,Kuhn:2005gv,Kallweit:2014xda,Kallweit:2015dum}. 

The QCD multijet processes are generated using \MGvATNLO at leading order (LO) in QCD with up to four additional partons in the matrix element calculations. The \ttbar and single top quark backgrounds are generated at NLO using \POWHEG~2.0 and 1.0, respectively~\cite{Nason:2004rx,Frixione:2007vw,powheg:2010,Nason:2013ydw,Alioli:2008gx,Alioli:2008tz}, 
and the diboson ($\PW\PW$, $\PW\PZ$, and $\PZ\PZ$) processes are generated at LO in QCD using 
\PYTHIA~8.205~\cite{Sjostrand:2014zea}.

The MC samples generated using \MGvATNLO and \POWHEG are interfaced to \PYTHIA~8.212~\cite{Sjostrand:2014zea} using the CUETP8M1 tune~\cite{Khachatryan:2015pea} for the fragmentation, hadronization, and underlying event. The \MGvATNLO generator provides jets from matrix-element calculations matched to the parton-shower description following the MLM~\cite{Mangano:2006rw} (FxFx~\cite{Frederix:2012ps}) method for LO (and NLO). We use the NNPDF~3.0~\cite{Ball:2014uwa} PDFs in all generated samples. All MC events are processed through a simulation of the CMS detector based on \GEANTfour~\cite{Agostinelli:2002hh} and are reconstructed with the same algorithms used for data. Additional pileup interactions are also simulated. The distribution of the 
number of pileup interactions in the simulation is adjusted to match the one observed in the data. The average number of pileup interactions in 2016 was 23.

\section{Signal extraction}
\label{sec:signal-extraction} 

The largest background contribution in this analysis, about 85\%, arises from the $\Wlvjets$ process in events where the charged lepton (electron, muon, or $\tau$-lepton) is either not reconstructed or falls outside the detector acceptance, and is estimated using data from the mutually exclusive CRs selected from single-muon and single-electron final states. The hadronic recoil \pt from the jets is used as an estimator for $\ptmiss$ in these control samples, and is defined by excluding any identified electrons or muons from its calculation.

A binned maximum-likelihood fit to the data is performed simultaneously in the SR and the two CRs to estimate the signal and the $\Wlvjets$ background rates in each $\ptmiss$ bin. The $\Wlvjets$ background normalization in each $\ptmiss$ bin is therefore a free parameter of the fit along with the signal yield:

\begin{equation*}\label{eqn:likeli}
  \begin{aligned}
  \mathcal{L}(\boldsymbol{\mu}^{\Wlv},\boldsymbol{\mu}, \boldsymbol{\theta}) &=
  \prod_{i} \mathrm{Poisson}\left(d_{i}     |B_{i}    (\boldsymbol{\theta}) + \mu^{\Wlv}_{i} + \boldsymbol{\mu} S_{i}(\boldsymbol{\theta})\right )\\
  &~\times \prod_{i}          \mathrm{Poisson}\left(d^{\Pe}_{i} |B^{\Pe}_{i}(\boldsymbol{\theta}) + \frac{ \mu^{\Wlv}_{i} }{R^{\Pe}_{i} (\boldsymbol{\theta})} \right)\\
  &~\times \prod_{i}          \mathrm{Poisson}\left(d^{\mu}_{i} |B^{\mu}_{i}(\boldsymbol{\theta}) + \frac{ \mu^{\Wlv}_{i} }{R^{\mu}_{i} (\boldsymbol{\theta})} \right).
\end{aligned}
\end{equation*}

In the above fit, the $d_{i}$, $d^{\Pe}_{i}$, and $d^{\mu}_{i}$ -- the symbol $i$ denotes each bin of the $\ptmiss$ distribution -- are the observed numbers of events in each bin of the SR, single-electron, and single-muon CRs, respectively. The expected contributions from other background processes in the CRs are denoted as $B_i^{\Pe}$ and $B_i^{\mu}$. The transfer factors, $R^{W}_{e}$ and $R^{W}_{\mu}$, are defined as the ratio of the expected yields of the target process in the signal region to the process being measured in the control sample; they are used to constrain the \Wlvjets backgrounds estimated in the SR. The parameter $\boldsymbol{\mu}_{i}^{\Wlv}$ represents the yield of the $\Wlvjets$ background in the $i^{th}$ $\ptmiss$ bin, and is left freely floating in the fit. The likelihood also includes a term for the signal region in which $B_i$ represents all the backgrounds apart from $\Wlvjets$, $S_i$ represents the nominal signal prediction, and $\mu$ denotes the observed cross section relative to the predicted value. The systematic uncertainties ($\boldsymbol{\theta}$) enter the likelihood as additive perturbations to the transfer factors, and are modeled in the maximum-likelihood fit with Gaussian priors as constrained nuisance parameters that include both experimental and theoretical components.

Both experimental and theoretical uncertainties affecting the transfer factors are included. Experimental uncertainties, including the lepton reconstruction, selection, and veto efficiencies are incorporated into the transfer factors. Lepton veto efficiencies are estimated by propagating them through the uncertainty in the tagging-scale factor to the vetoed selections, whereas the lepton flavor composition of the \Wlvjets process is constrained by the CRs. These experimental uncertainties are only applicable to the leptons in the SR that are not identified, and therefore passed the veto requirement but fall into the detector acceptance. The overall magnitude of the lepton veto uncertainty is determined to be $\approx$2\% and is dominated by the $\tau$ lepton veto uncertainty. The uncertainties in the efficiency of the electron and $\ptmiss$ triggers are also included in the transfer factors. Theoretical uncertainties of the \Wlvjets contributions include the effects from the modeling of uncertainties in the PDFs~\cite{Butterworth:2015oua} due to higher-order corrections. All uncertainties discussed are within 1--2\% range and are fully correlated across all the bins of hadronic recoil \pt. The full list of uncertainties in the transfer factors are summarized in Table~\ref{tab:systematics}.

\begin{table}[htbp]
\centering
\topcaption{\label{tab:systematics}
  Experimental uncertainties affecting transfer factors in the analysis that is used to estimate the $\Wlv$ background in the SR. 
  The number of $\PW$ boson events are denoted as $\PW_{\mathrm{SR}}$ for the SR and in analogy as $\PW_{\PGm\PGn}$ ($\PW_{\Pe\PGn}$) for the single-muon (single-electron) CR.
  The term shape describes the variation in the uncertainty across the hadronic recoil $\pt$ or $\ptmiss$ spectra.}
\begin{tabular}{llc}
\hline
Source  		           & Process                            & Uncertainty (\%) \\
\hline
Electron trigger		   & $\PW_{\mathrm{SR}}/\PW_{\Pe\PGn}$            & 1 \\
$\ptmiss$ trigger                  & $\PW_{\mathrm{SR}}/\PW_{\Pe\PGn/\PGm\PGn}$   & 0--2 (shape) \\
Muon reconstruction efficiency     & $\PW_{\mathrm{SR}}/\PW_{\PGm\PGn}$           & 1 \\
Muon identification efficiency     & $\PW_{\mathrm{SR}}/\PW_{\PGm\PGn}$           & 1 \\
Electron reconstruction efficiency & $\PW_{\mathrm{SR}}/\PW_{\Pe\PGn}$            & 1 \\
Electron identification efficiency & $\PW_{\mathrm{SR}}/\PW_{\Pe\PGn}$            & 1.5 \\
Muon veto			   & $\PW_{\mathrm{SR}}/\PW_{\Pe\PGn/\PGm\PGn}$   & $<$1 (shape) \\
Electron veto			   & $\PW_{\mathrm{SR}}/\PW_{\Pe\PGn/\PGm\PGn}$   & $<$1 (shape) \\
$\tau$ veto			   & $\PW_{\mathrm{SR}}/\PW_{\Pe\PGn/\PGm\PGn}$   & 1--2 (shape) \\
PDF				   & $\PW_{\mathrm{SR}}/\PW_{\Pe\PGn/\PGm\PGn}$   & 1--2 (shape) \\
\hline
\end{tabular}
\end{table}

The remaining backgrounds that contribute to the total event yield in the SR are much smaller than those from $\Wlvjets$. 
These backgrounds include QCD multijet events, that are measured by extrapolating data from the CR defined by $\Delta\phi(\ptvecjet,\ptvecmiss)<0.5$. Top quark, diboson, and $\Zlljets$ (when both leptons are out of acceptance) production are obtained directly from simulation.

Uncertainties assigned to processes estimated from simulation include the uncertainties in the efficiency of the \PQb jet veto, which are estimated to be 3.0 (1.0)\% for the top quark (diboson) background. A systematic uncertainty of 10\% is included in the top quark background normalization from the modeling of the top quark $\pt$ distribution~\cite{Czakon:2017wor,Khachatryan:2016mnb}. Systematic uncertainties of 10 and 20\% are included in the normalizations of the top  quark~\cite{Khachatryan:2015uqb} and diboson 
backgrounds~\cite{Khachatryan:2016txa,Khachatryan:2016tgp}, respectively, to account for the uncertainties in their cross sections in the relevant kinematic phase space. A systematic uncertainty of 20\% is also used to account for the rate of the $\Zlljets$ process, which constitutes less than 0.4\% of the total background since this process can only arise in events where the charged lepton is either not reconstructed or falls outside the detector acceptance. The uncertainty in the measurement of the integrated luminosity of 2.5\%~\cite{CMS-PAS-LUM-17-001} is assigned to all simulation-based processes. The jet energy scale uncertainties are less than 3\% for jets within the tracker acceptance and reach up to 6\% for those outside. The uncertainty in the momenta of jets is then propagated to $\ptmiss$ by varying the momenta of each jet within its uncertainty and recomputing $\ptmiss$~\cite{Sirunyan:2019kia}. The resulting uncertainty due to jet/$\ptmiss$ momentum scale is calculated to be 4\%. Finally, an uncertainty in the QCD multijet background of 50--150\% is assigned to cover differences in the jet response and the statistical uncertainty in the extrapolation from the CR to the SR. These uncertainties are summarized in Table~\ref{tab:systematics_backgrounds}.

A cross-check unfolding method was performed, where the background contributions were estimated through a likelihood fit and were subtracted from the data. The alternative unfolding procedure consists of performing a least-squares fit with optional Tikhonov regularization \cite{Tikhonov:1963}, as implemented in the TUnfold software package \cite{Schmitt:2012kp}. The results from this method agree very well with the standard procedure.

\begin{table}[htbp]
\centering
\topcaption{\label{tab:systematics_backgrounds}
  Uncertainties assigned to processes in the SR and CRs. The term shape describes the variation in the uncertainty across the hadronic recoil $\pt$ or $\ptmiss$ spectra.}
\cmsTable{
\begin{tabular}{llc}
\hline
Source                             & Process	                                & Uncertainty (\%)\\
\hline
Luminosity~\cite{CMS-PAS-LUM-17-001}		           & All        		      & 2.5 \\
Electron trigger	           & All in single-electron CR        & 1.0 \\
$\ptmiss$ trigger	           & All in SR and single-muon CR     & 0--2.0 (shape) \\
Jet/$\ptmiss$ momentum scale       & All        		      & 4.0 \\
Pileup  		           & All        		      & 1.0--2.0 (shape) \\
Muon reconstruction efficiency     & All in single-muon CR	      & 1.0 \\
Muon identification efficiency     & All in single-electron CR        & 1.0 \\
Electron reconstruction efficiency & All in single-electron CR        & 1.0 \\
Electron identification efficiency & All in single-electron CR        & 1.5 \\
\PQb jet veto		           & Top quark in SR and all CRs      & 3.0 \\
			           & All remaining in SR and all CRs  & 1.0 \\
$\pt$ reweighting of top quark     & Top quark 		              & 10 \\
Top quark normalization	           & Top quark		              & 10 \\
Normalization of diboson           & Diboson		              & 20 \\
$\Zlljets$ normalization           & \Zlljets in SR	              & 20 \\
Normalization of QCD multijets     & QCD multijets in SR              & 50--150 (shape) \\
\hline
\end{tabular}
}
\end{table}

\section{Fiducial cross section measurements}
\label{sec:cross_section}

This analysis provides the \PZ boson production cross section in a restricted fiducial region and compares with the theoretical predictions. The data selection defined at the reconstruction level is summarized in Table~\ref{tab:selection}. The fiducial phase space, for which theoretical predictions are computed, is 
defined at the generator level without considering the detector response. The detector resolution leads to a difference between the observables at the reconstruction and the analogous generator level quantities. To minimize discrepancies, the selection criteria imposed at the generator level are defined to mimic the definition at the reconstructed level as much as possible. For this analysis, the fiducial phase space is defined by requiring $\ptmiss>200\GeV$  without any other requirements placed on the \PZ boson kinematics or the transverse momentum of the leading jet. The $\ptmiss$ is the reconstructed $\pt$ of the neutrino pair emitted in \PZ boson decays. 

The difference between fiducial and analysis requirements at the reconstructed level, specifically the requirements on $\ptmiss>250\GeV$ and $\pt^{\text{jet}}>100\GeV$, introduces a small, $\approx$0.9\%, theoretical uncertainty from PDFs and missing higher orders in renormalization and factorization scales (QCD scales). These uncertainties are computed by comparing the ratio of theoretical uncertainties in the tighter to the looser analysis phase space requirements. Finally, the contribution arising from events reconstructed within the fiducial phase space, but not originating from there, is treated as background.

To measure the total and differential production cross sections, events in the SR are divided into ten bins of $\ptmiss$, where each bin width is chosen to be at least equal to the $\ptmiss$ resolution at the given $\ptmiss$ range. A detailed description of the performance of the $\ptmiss$ observable is presented in Ref.~\cite{Sirunyan:2019kia} using $\Zjets$ and $\gamma$+jets events. Based on these $\ptmiss$ measurements, the minimum bin width is chosen to be 25\GeV at the lower end of the spectrum and is gradually increased to 350\GeV at the highest $\ptmiss$. The width of the highest $\ptmiss$ bin is also chosen to reflect the statistical precision of the sample. 

\begin{figure}[htbp]
\centering
\includegraphics[width=0.46\textwidth]{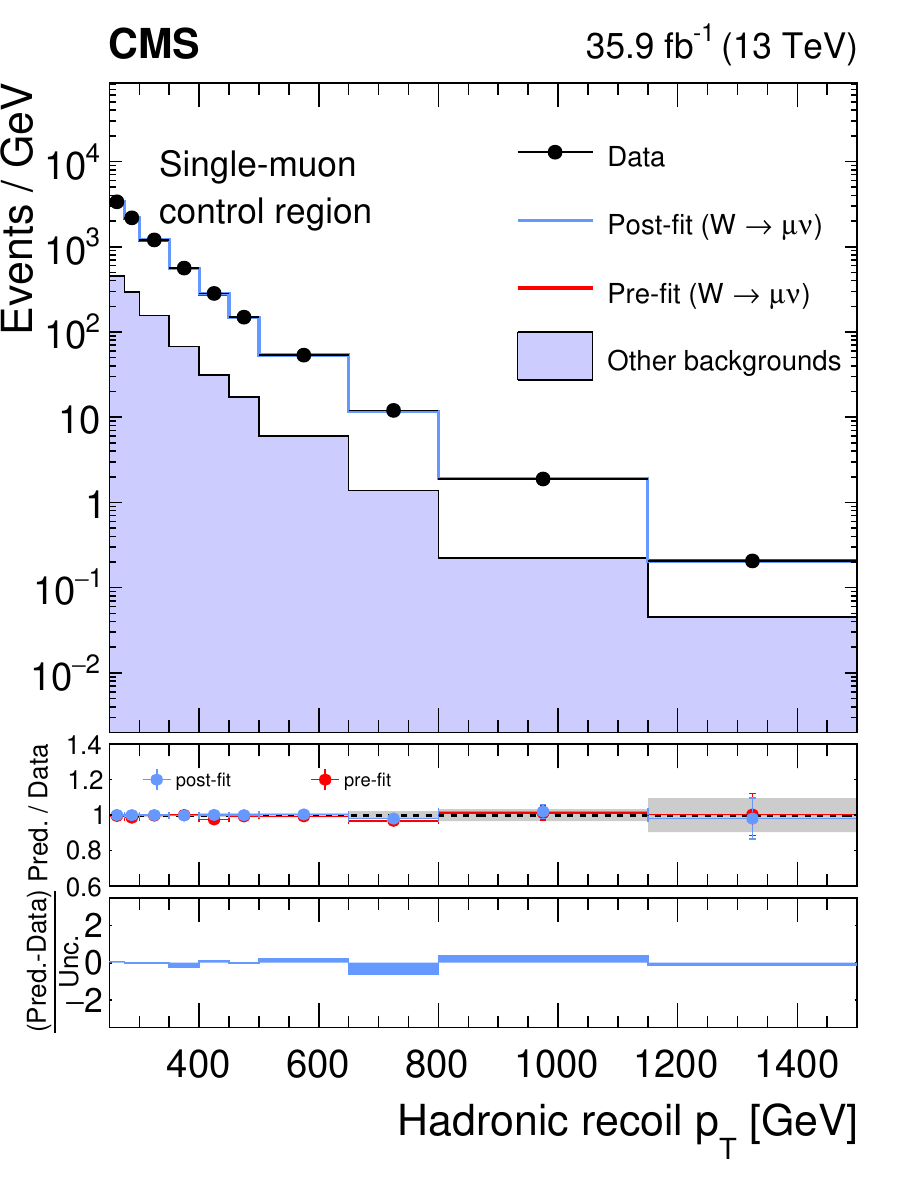}
\includegraphics[width=0.46\textwidth]{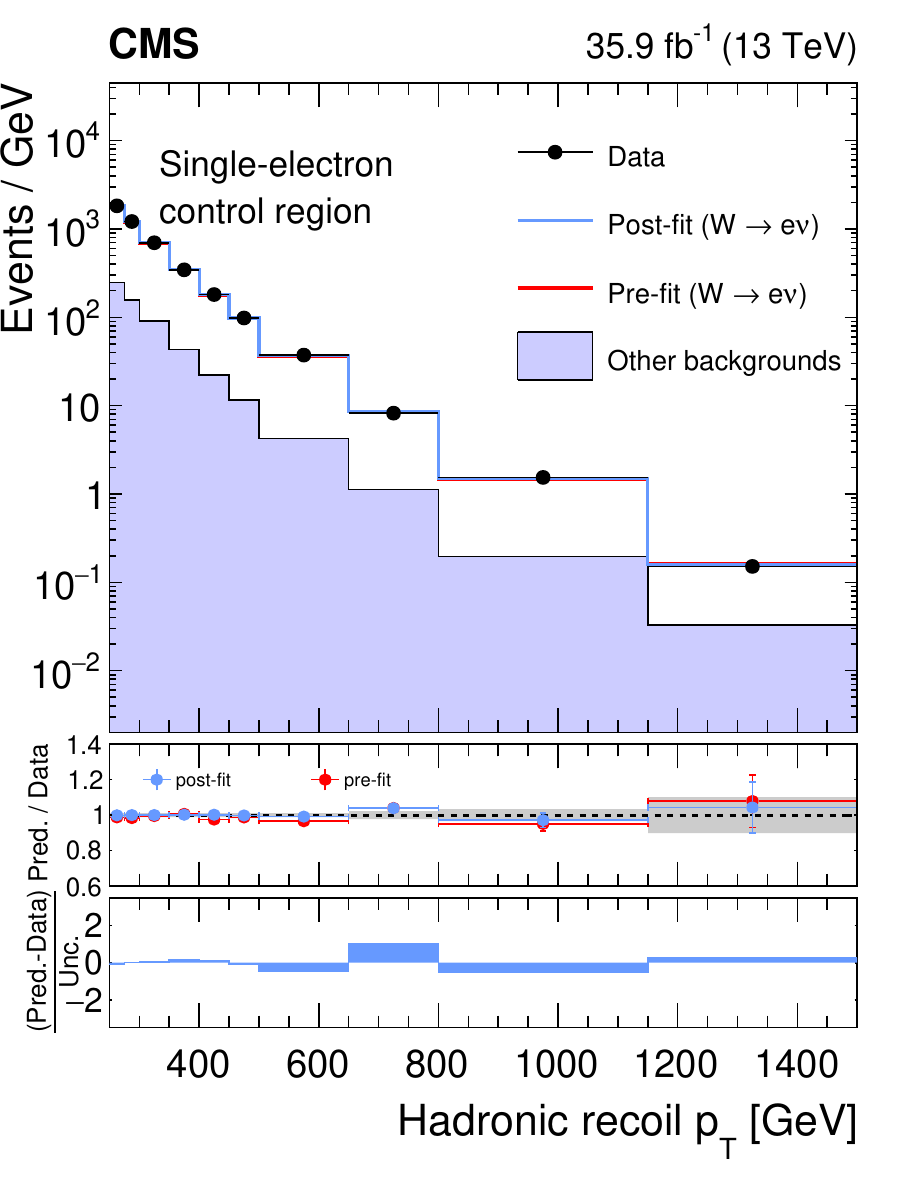}\\
\includegraphics[width=0.46\textwidth]{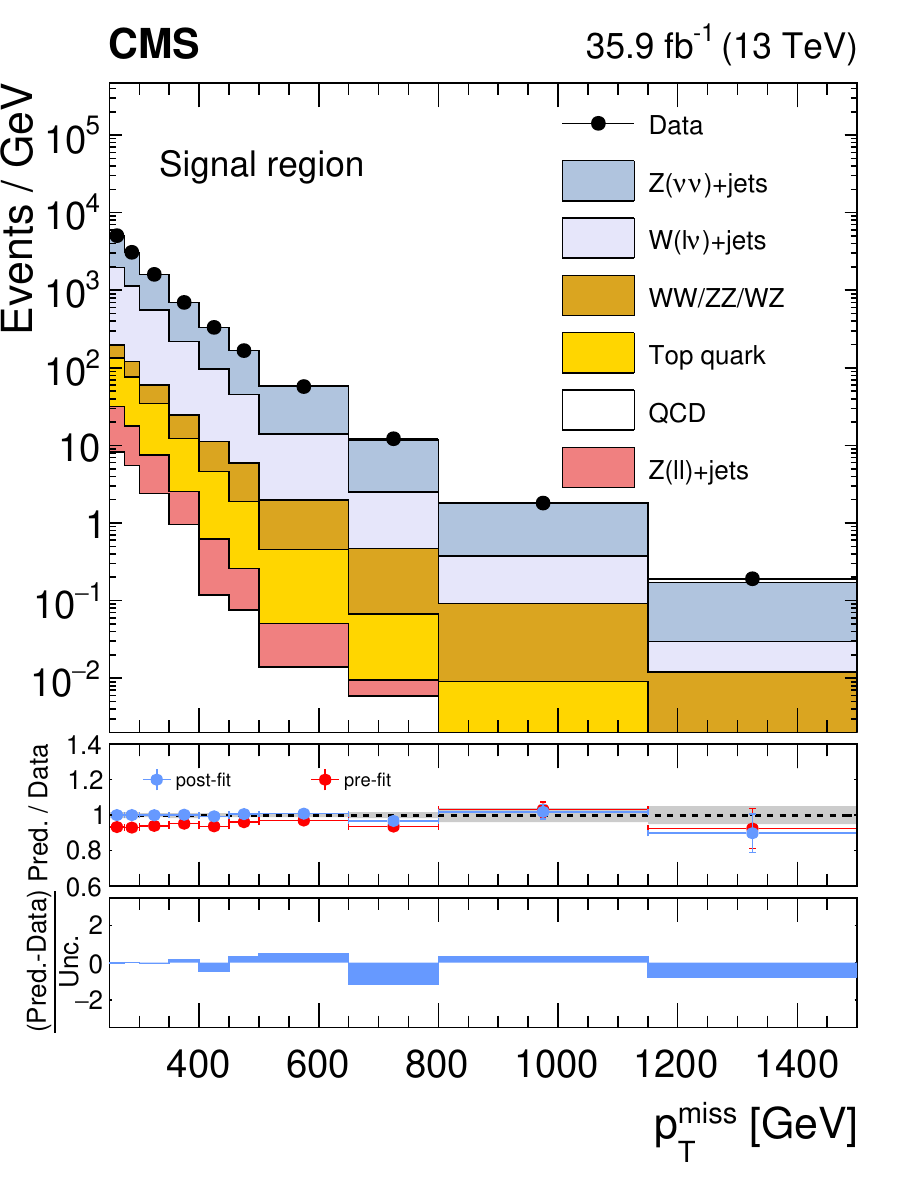}
\caption{\label{fig:fits_control}
  Comparison of data and simulation in the single-muon (upper left), single-electron (upper right) CRs and in the SR (lower), before and after performing the simultaneous fit across all the signal and control regions. The hadronic recoil $\pt$ in single lepton events is used as an estimator for $\ptmiss$ in the SR. For the distributions in the CRs, the other backgrounds include top quark, diboson, and QCD multijet events. The post-fit distributions are shown in the upper panel. Ratios of data with the pre-fit expectation (red points) and post-fit prediction (blue points) are shown. The gray band in the ratio panel indicates the post-fit uncertainty after combining all systematic uncertainties. The distribution of the pulls, defined as the difference between data and the post-fit expectation relative to the quadratic sum of the post-fit uncertainties in the expectation, and statistical uncertainty in data, are shown in the lower panel.}
\end{figure}

As explained above, to extract the signal and the background, a maximum-likelihood fit is performed using the SR and the CRs. Figure~\ref{fig:fits_control} shows the comparison between data and expectations in the SR and CRs before (pre-fit) and after (post-fit) the fit to the data. Pre-fit distributions for the $\Zjets$ and $\Wjets$ processes are based on simulated MC samples produced at NLO in QCD and are corrected using vector boson $\pt$-dependent higher-order EW corrections extracted from theoretical calculations. We verified that the measured cross sections are not sensitive to these corrections.

\begin{figure}[htbp]
\centering
\includegraphics[width=0.49\textwidth]{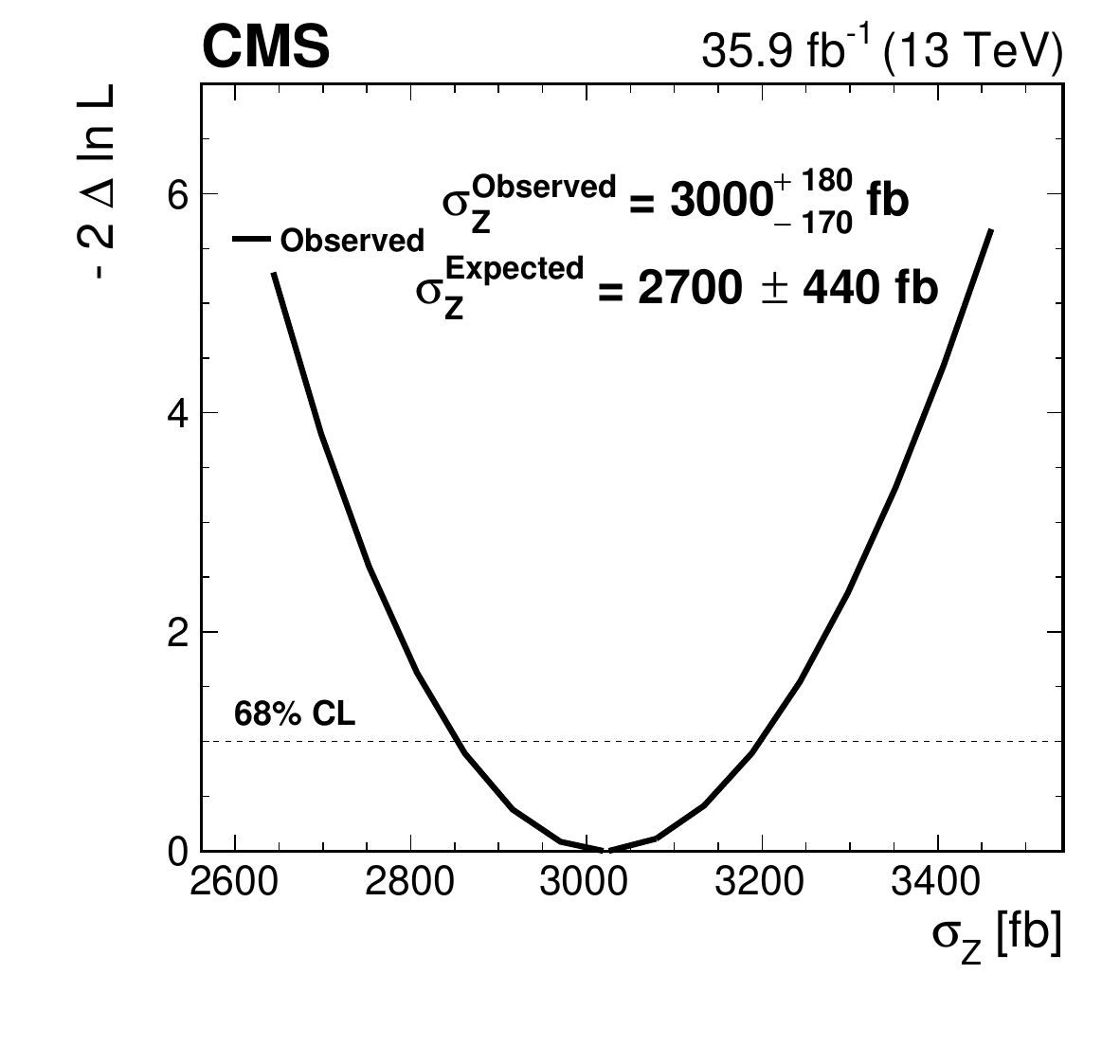}
\caption{\label{fig:xsec}
  The likelihood scan for the fiducial \PZ boson production cross section in the $\PZ \to \PGn\PGn$ channel.}
\end{figure}

The measured total fiducial cross section is $3000^{+180}_{-170}$\unit{fb}. The corresponding likelihood scan is shown in Fig.~\ref{fig:xsec}, together with the predicted value $\sigma_{\PZ\to\PGn\PGn} = 2700\pm440$\unit{fb} from \MGvATNLO using the NNPDF 3.0~\cite{Ball:2014uwa} NLO PDFs. The theoretical uncertainties for \MGvATNLO include statistical, PDF, and QCD-scale uncertainties. The PDF uncertainties are estimated  by taking the one standard deviation band of the predictions from the replicas available for samples~\cite{Ball:2017nwa}. The QCD-scale uncertainties are estimated by changing the renormalization ($\mu_{\mathrm{R}}$) and factorization ($\mu_{\mathrm{F}}$) scales independently up and down by a factor of two from their nominal values 
(excluding the two extreme values) and taking the largest changes in the cross section as the uncertainty. 
The dominant experimental uncertainties are associated with the jet and $\ptmiss$ momentum scales and the integrated luminosity.

\begin{figure}[htbp]
\centering
\includegraphics[width=0.46\textwidth]{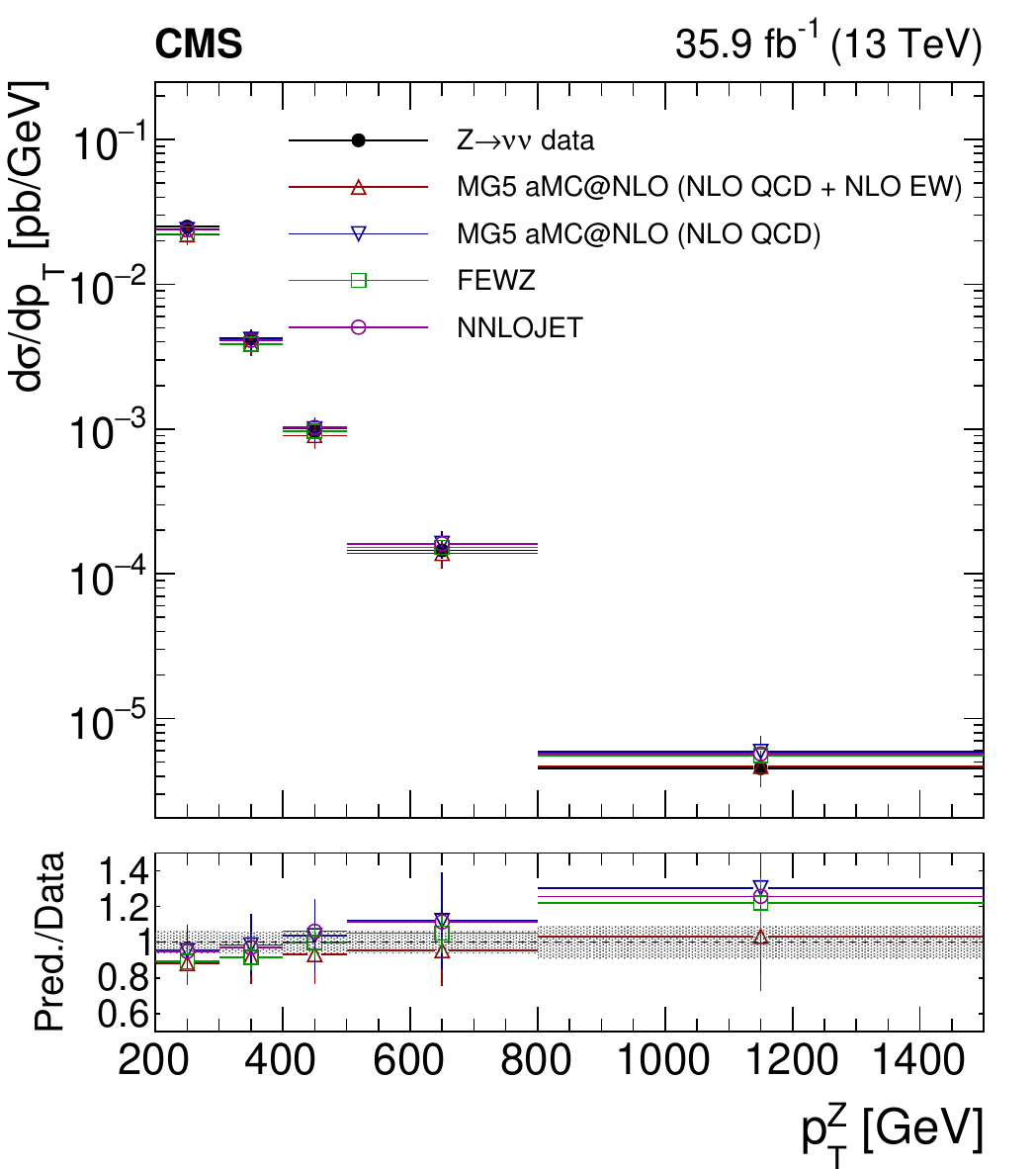}
\includegraphics[width=0.46\textwidth]{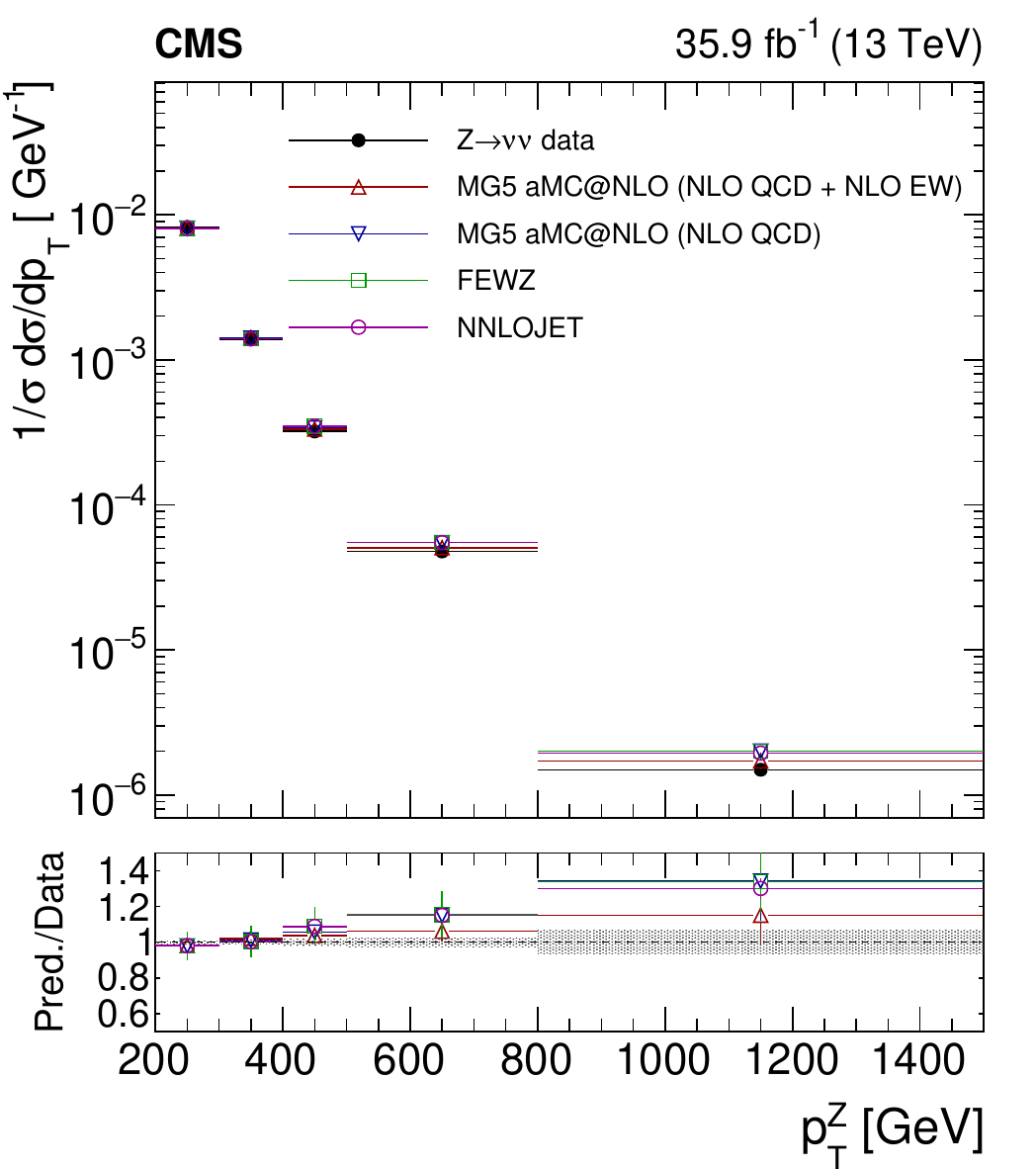}
\caption{\label{fig:xsec_diff}
  The measured absolute (left) and normalized (right) fiducial cross sections as a function of $\zpt$ compared with \MGvATNLO and fixed-order calculations. The shaded bands around the data points correspond to the total experimental uncertainty. The vertical bars around the predictions correspond to the combined statistical, PDF, and QCD-scale uncertainties.
 }
\end{figure}

In the differential cross section measurement, the signal templates are determined from five differential fiducial regions in $\ptmiss$, and each is allowed to float in the binned fit. The fiducial regions are obtained by combining two neighboring ranges in the reconstructed $\ptmiss$ spectra. This stabilizes the fit and ensures that at least 65\% of the events in the reconstructed $\ptmiss$ bin are also in the same bin in terms of the true $\pt$ of the decaying \PZ boson. The differential cross section measurements as functions of $\zpt$ are shown in Fig.~\ref{fig:xsec_diff} (left) with predictions from \MGvATNLO 
at NLO in QCD and with or without the higher-order EW corrections. These corrections are important at high \zpt with expected correction factors of $\approx$0.9 at $\zpt=500\GeV$ and $\approx$0.8 at $\zpt = 1000\GeV$~\cite{Dittmaier:2014qza,Lindert:2017olm}.

The cross section as a function of $\zpt$ distribution is also compared to the fixed-order expectation from $\FEWZ$ 3.1~\cite{FEWZ, Gavin:2010az, Gavin:2012sy, Li:2012wna} at NNLO accuracy in QCD ($\mathcal{O}(\alpS^2)$, where $\alpS$ is the strong coupling) using the NNPDF 3.1~\cite{Ball:2017nwa} NNLO PDFs. The central values of $\mu_{\mathrm{F}}$ and $\mu_{\mathrm{R}}$ are chosen to be $\sqrt{\smash[b]{q_{\mathrm{T}}^{2}+m_{\PZ}^{2}}}$, where $m_{\PZ}$ is the nominal \PZ boson mass~\cite{PDG2020} and $q_{\mathrm{T}}$ is the value of the lower edge of the corresponding bin in $\pt^{\PZ}$. 
The uncertainties in $\FEWZ$ include statistical, PDF, and QCD-scale uncertainties, calculated the same way as for \MGvATNLO.

Finally, the data are compared with the NNLO predictions of vector boson production in association with a jet~\cite{Ridder:2015dxa} at $\mathcal{O}(\alpS^3)$ accuracy using the NNPDF 3.1~\cite{Ball:2017nwa} NNLO PDFs (referred as NNLOJET in the figures). The central values of $\mu_{\mathrm{F}}$ and $\mu_{\mathrm{R}}$ are chosen to be $\sqrt{\smash[b]{(\pt^{\PZ})^2+\mll^{2}}}$. 
The uncertainties for $\PZ$+1 jet at NNLO include PDF and QCD-scales, calculated the same way as for \MGvATNLO, and they are largely reduced with respect to other calculations.

Similar to the inclusive measurement, the dominant source of uncertainties are associated with the jet and $\ptmiss$ scales and the integrated luminosity. Whereas the systematic uncertainties dominate in the first four fiducial regions, statistical uncertainty dominates in the last signal bin.

The ratio of the differential cross section relative to the total fiducial cross section (normalized cross section) is also measured using the same binning. In this measurement, the total and individual cross sections are evaluated simultaneously, and therefore, the systematic uncertainties in the individual cross sections are largely reduced. 
While each fiducial bin is measured separately, only four of these contributions are used to float in the binned maximum-likelihood fit, along with their sum. In this way, the normalized cross section has the same number of degrees of freedom as the differential measurement. The differential cross section measurements normalized to the total cross section are presented in Fig.~\ref{fig:xsec_diff} (right), as well as cross section ratios with respect to the predicted $\zpt$ measurements.

\section{Combination of cross section measurements using charged leptons and neutrinos}
\label{sec:combination}

Measurements of differential \PZ boson production cross section have also been performed with \PZ bosons decaying to dielectrons or dimuons~\cite{CMS_Z_13TeV}, and these are combined with our new  measurements to improve the precision at large $\zpt$. In both analyses, the signal samples are generated with \MGvATNLO implemented with the NLO EW corrections applied~\cite{Lindert:2017olm}.

While the analysis selection is identical, the fiducial region of the charged-lepton case is modified compared to Ref.~\cite{CMS_Z_13TeV}, to match the definition in Section~\ref{sec:cross_section}.  Specifically, the requirement on the dilepton mass to lie within 15\GeV of the \PZ boson mass is kept to reduce the photon propagator contribution, but no requirements on the generator-level $\pt^\ell$ and $\eta^\ell$ are applied. The removal of the fiducial requirements on kinematic observables of the leptons is studied in simulation and introduces a small, $\approx$2\%, theoretical extrapolation uncertainty from PDF and QCD-scale changes in the $\PZ \to \ell\ell$ channel because the reconstruction-level selection requires $\pt^\ell>25\GeV$ and $\abs{\eta^\ell}<2.4$.

The leading systematic uncertainties between the two analyses are rather different. For the final state with neutrinos, the jet and $\ptmiss$ momentum scales uncertainties are dominant, whereas for the charged-lepton final states they originate from lepton identification~\cite{CMS_Z_13TeV}. Both the jet and $\ptmiss$ momentum scales uncertainties and the uncertainties on the lepton identification are kept uncorrelated between the two analyses. The only correlated uncertainty is the integrated luminosity. 

The signal cross section is extracted through a simultaneous binned maximum-likelihood fit to the signal and background $\ptmiss$ spectra in the SR and CRs of the neutrino channel, as described in Section~\ref{sec:cross_section}, and to the $\pt^{\ell\ell}$ spectra in the SRs of the charged lepton channel. The individual analyses and the combined differential cross sections are summarized in Table~\ref{tab:xs_zxxpt}.

\begin{table}[htbp]
\centering
\topcaption{\label{tab:xs_zxxpt}
  Cross sections (fb) at large $\zpt$ values in the $\PZ \to \ell\ell$ and $\PZ \to \PGn\PGn$ channels, and their combination. The theoretical predictions from $\MGvATNLO$ at NLO in QCD and corrected to NLO in EW~\cite{Lindert:2017olm} using the NNPDF~3.0 are also reported. With the exception of the largest $\zpt$ bin, the statistical uncertainties in the measurements are much smaller than the systematic  uncertainties. The measurements and predictions correspond to $\sigma \mathcal{B}(\PZ \to \ell\ell)$, where $\sigma$ is the total fiducial cross section, $\mathcal{B}$ is the branching fraction, and $\ell$ is a charged lepton representing electrons and muons. The $\PZ \to \nu\nu$ measurement corresponds to $\sigma \mathcal{B}(\PZ \to \ell\ell)/\mathcal{B}(\PZ \to \nu\nu)$.}
\begin{tabular}{lcccccc}
\hline
$\zpt$ (\GeVns{}) & $\Zee$ & $\Zmm$ & $\Zll$ & $\Zvv$ & $\Zllvv$ & Theory \\ \hline
200--300  & $2500^{+140}_{-110}$ & $2400^{+120}_{-120}$ & $2500^{+100}_{-100}$ & $2500^{+150}_{-150}$ & $2500^{+82}_{-100}$  & $2200 \pm 350$ \\ [\cmsTabSkip]
300--400  & $390^{+22}_{-18}$    & $400^{+22}_{-21}$    & $400^{+17}_{-16}$    & $420^{+24}_{-23}$    & $410^{+14}_{-17}$    & $390  \pm 69$ \\ [\cmsTabSkip]
400--500  & $99^{+5.7}_{-4.9}$   & $97^{+6.4}_{-6.1}$   & $100^{+4.4}_{-4.2}$  & $97^{+5.6}_{-5.4}$   & $97^{+3.3}_{-4.0}$   & $90   \pm 18 $ \\ [\cmsTabSkip]
500--800  & $47^{+3.0}_{-2.5}$   & $41^{+4.0}_{-3.7}$   & $45^{+2.3}_{-2.2}$   & $44^{+2.7}_{-2.6}$   & $44^{+1.6}_{-1.9}$   & $41   \pm 9.0$ \\ [\cmsTabSkip]
800--1500 & $3.9^{+0.6}_{-0.5}$  & $3.2^{+0.7}_{-0.6}$  & $3.7^{+0.4}_{-0.4}$  & $3.2^{+0.3}_{-0.3}$  & $3.3^{+0.2}_{-0.2}$  & $3.3  \pm 0.9$ \\ [\cmsTabSkip]
200--1500 & $3000^{+160}_{-130}$ & $3000^{+150}_{-140}$ & $3000^{+120}_{-110}$ & $3000^{+180}_{-170}$ & $3000^{+100}_{-120}$ & $2700 \pm 440$\\ [\cmsTabSkip]
\hline 
\end{tabular}
\end{table}

The combined result leads to a reduction in uncertainties compared with either of the two channels. In the lower-$\zpt$ regime, the combination is systematically limited and dominated by the charged-lepton channels. At the same time, at higher $\zpt$, the statistical limitation of the charged-lepton channel is mitigated by the $\Zvv$~channel, yielding improved sensitivity. 

\begin{figure}[htbp]
  \centering
  \includegraphics[width=0.46\textwidth]{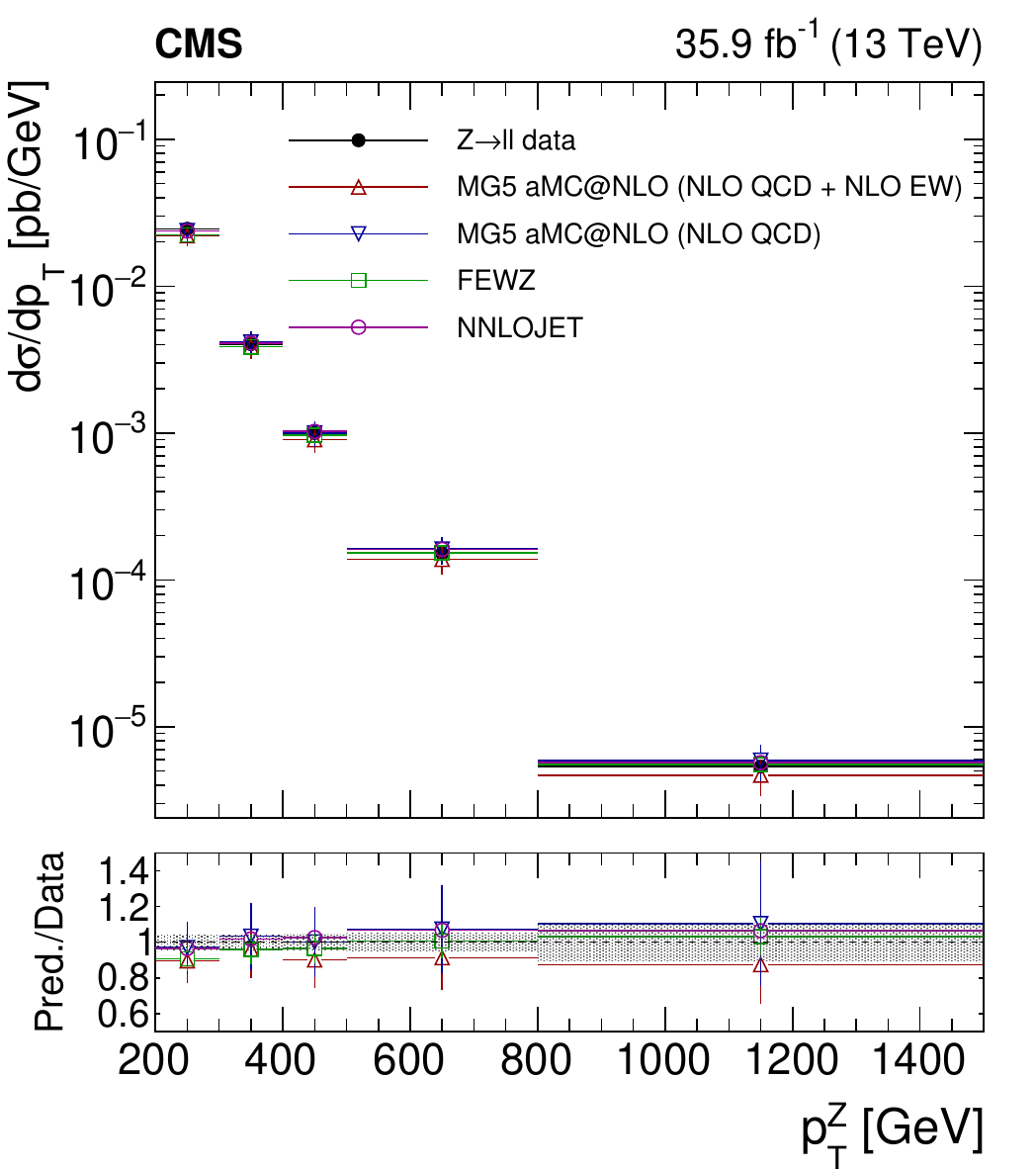}
  \includegraphics[width=0.46\textwidth]{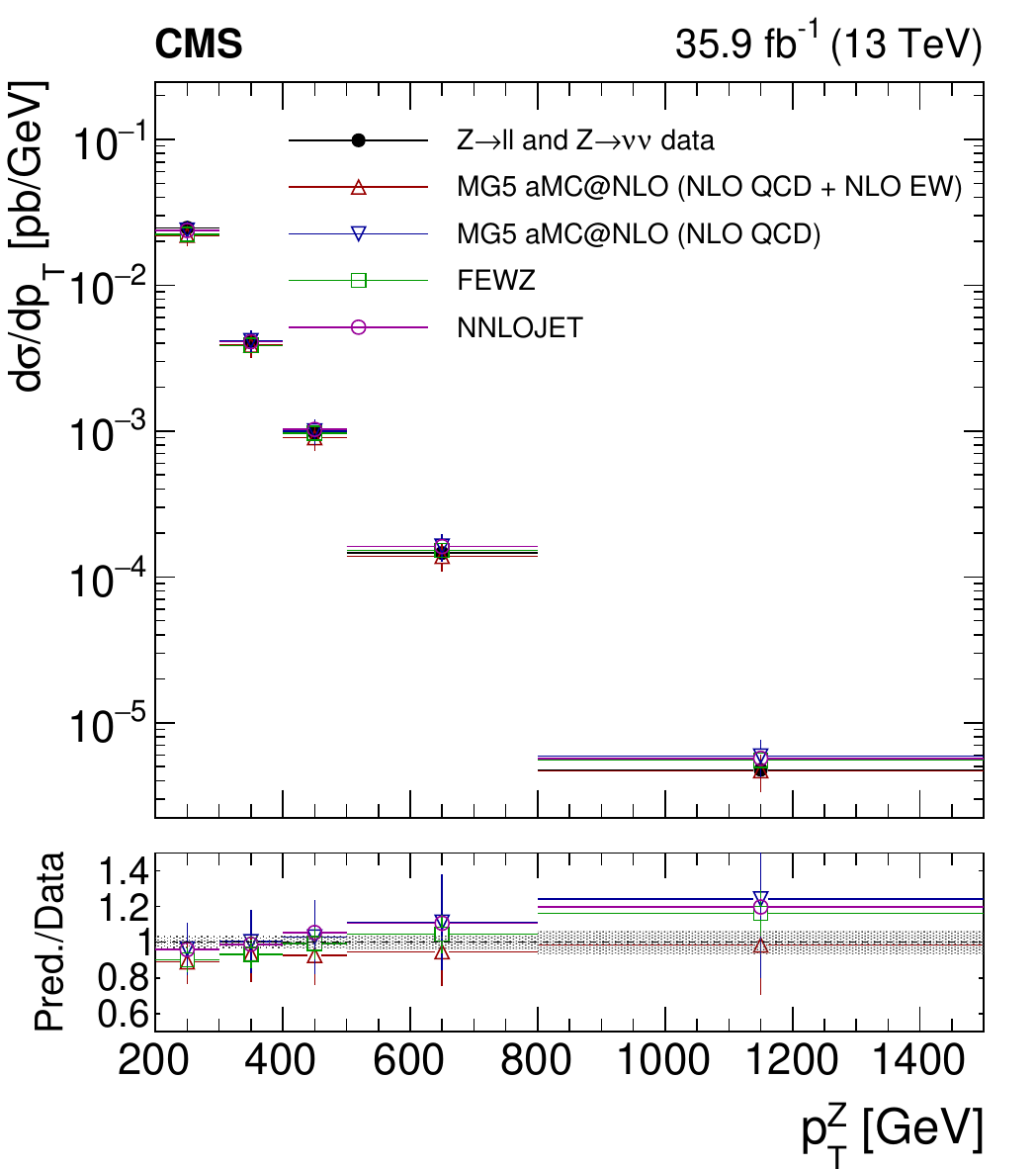}
  \caption{\label{fig:unf_pt}
    Measured $\zpt$ absolute cross section for $\PZ\to\ell^+\ell^-$ (left), and the combination (right) being compared with \MGvATNLO and fixed-order calculations. The shaded bands around the data points correspond to the total experimental uncertainty. The vertical bars around the predictions correspond to the combined statistical, PDF, and QCD-scale uncertainties.}
\end{figure}

The measured experimental distributions are compared with the theoretical predictions from the \MGvATNLO generator with and without NLO EW corrections. These distributions are shown in Fig.~\ref{fig:unf_pt}. The uncertainty in the theoretical predictions includes uncertainties both from PDF and from renormalization and factorization scales, together with the statistical precision of the available samples.

\begin{figure}[htbp]
  \centering
  \includegraphics[width=0.46\textwidth]{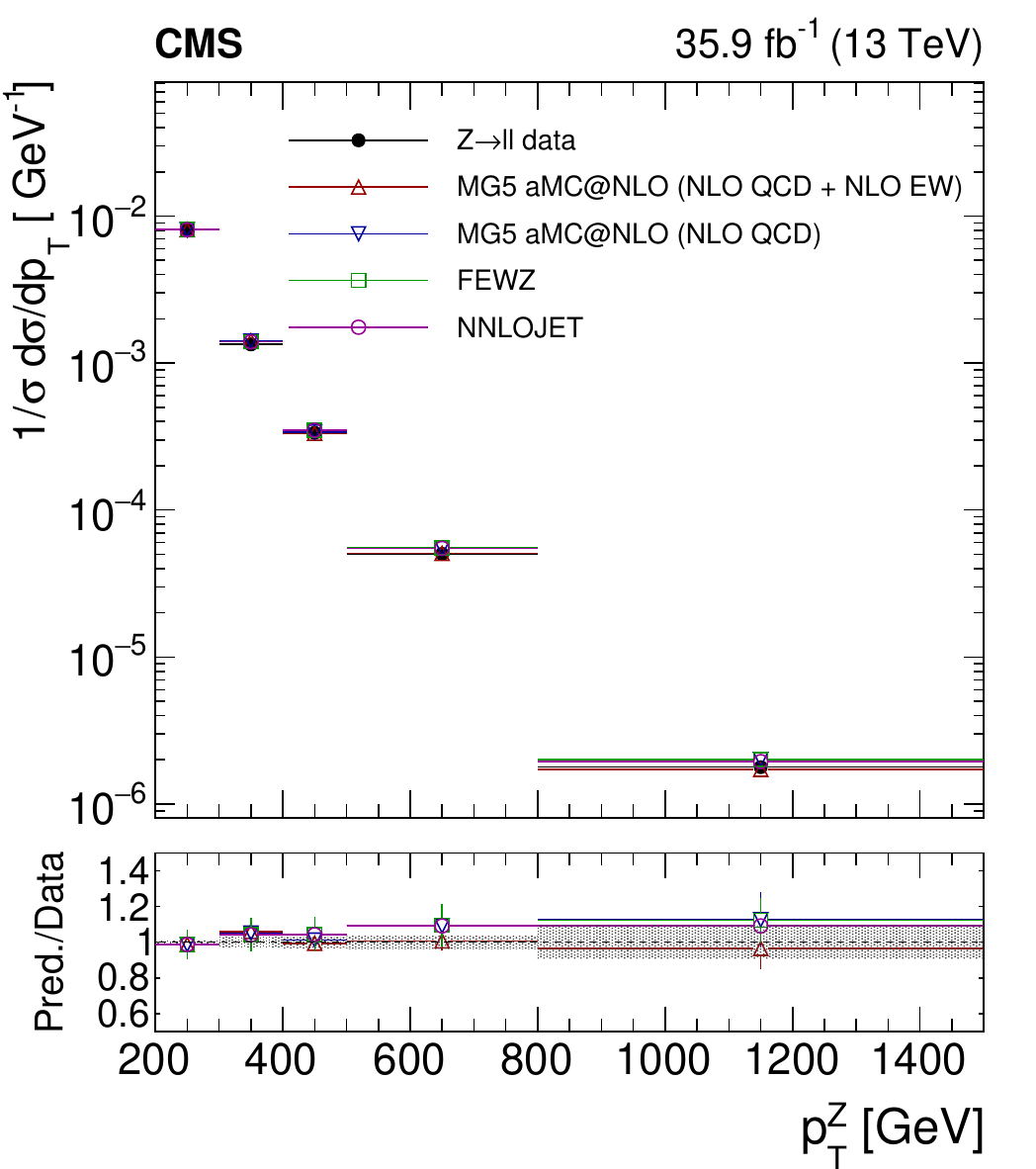}
  \includegraphics[width=0.46\textwidth]{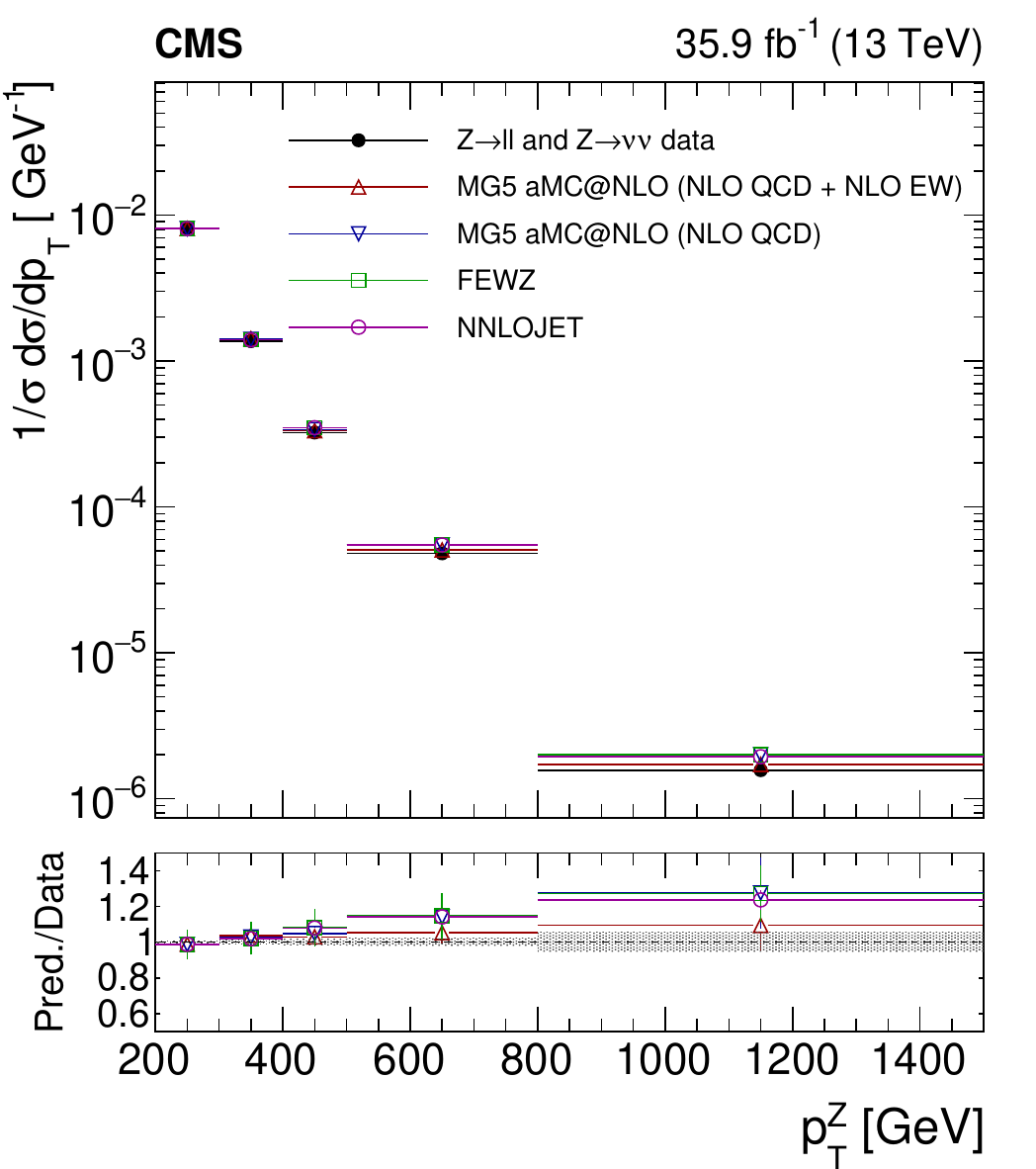}
  \caption{\label{fig:unf_norm_pt}
    Measured $\zpt$ normalized cross section for $\PZ\to\ell^+\ell^-$ (left), and the combination (right) compared with \MGvATNLO and fixed-order calculations. The shaded bands correspond to the total systematic uncertainty. The vertical bars around the predictions correspond to the combined statistical, PDF, and QCD-scale uncertainties.}
\end{figure}

The combined cross sections normalized to the total cross section are presented in Fig.~\ref{fig:unf_norm_pt}. Table~\ref{tab:xs_zxxpt_normalized} presents the normalized cross sections relative to the predicted normalized values. The uncertainties due to the jet and $\ptmiss$ momentum scales have a smaller $\zpt$ distribution dependence than the lepton efficiency uncertainties. Therefore, by evaluating the ratio of cross sections, the uncertainties in the $\Zvv$ channel are more reduced than the ones from the $\Zll$ channels. While predictions are consistent with data within the experimental and theoretical uncertainties in the full range of $\zpt$, a deviation up to 15\% at the highest-$\zpt$ regime is observed.

\begin{table}[htbp]
\centering
\topcaption{\label{tab:xs_zxxpt_normalized}
Cross sections normalized to the total cross section measurements relative to 
the predicted values in the $\PZ\to\ell\ell$ and $\PZ\to\PGn\PGn$ channels, 
and in their combination. The predicted values are reported in Table~\ref{tab:xs_zxxpt}, 
and the measured $\zpt$ normalized cross sections are shown in Fig.~\ref{fig:unf_norm_pt}.}
\begin{tabular}{lccc}
\hline 

$\zpt$ (\GeVns{}) & $\Zll$ & $\Zvv$ & $\Zllvv$\\ \hline
200--300  & $1.012^{+0.006}_{-0.007}$ & $1.019^{+0.009}_{-0.009}$ & $1.011^{+0.004}_{-0.004}$ \\ [\cmsTabSkip]
300--400  & $0.943^{+0.025}_{-0.025}$ & $0.979^{+0.015}_{-0.015}$ & $0.963^{+0.015}_{-0.014}$ \\ [\cmsTabSkip]
400--500  & $1.006^{+0.031}_{-0.030}$ & $0.963^{+0.019}_{-0.019}$ & $0.971^{+0.017}_{-0.016}$ \\ [\cmsTabSkip]
500--800  & $0.993^{+0.036}_{-0.036}$ & $0.942^{+0.024}_{-0.024}$ & $0.949^{+0.021}_{-0.020}$ \\ [\cmsTabSkip]
800--1500 & $1.036^{+0.099}_{-0.095}$ & $0.869^{+0.059}_{-0.057}$ & $0.914^{+0.052}_{-0.051}$ \\ [\cmsTabSkip]
\hline 
\end{tabular}
\end{table}

\section{Summary}
\label{sec:summary}

Total, differential, and normalized fiducial cross section measurements for a \PZ boson produced in association with one or more jets in proton-proton collisions at a center-of-mass energy of $13\TeV$ at high \PZ boson $\pt$ in the invisible decay channel ($\Zvvbar$) have been presented. The data collected with the CMS detector at the LHC correspond to an integrated luminosity of 35.9\fbinv. The measured total fiducial cross section for events with \PZ boson transverse momentum greater than 200\GeV is $3000^{+180}_{-170}$\unit{fb}. The precision of this result is improved by combining the cross section measured with those extracted from charged-lepton final states. The results agree within uncertainties with the theoretical predictions from \MGvATNLO, $\FEWZ$ and $\PZ$+1 jet at next-to-next-to-leading order in perturbative quantum chromodynamics. These are the most precise measurements of the $\zpt$ spectrum to date in proton-proton collisions at 13\TeV.

\begin{acknowledgments}
  We congratulate our colleagues in the CERN accelerator departments for the excellent performance of the LHC and thank the technical and administrative staffs at CERN and at other CMS institutes for their contributions to the success of the CMS effort. In addition, we gratefully acknowledge the computing centers and personnel of the Worldwide LHC Computing Grid for delivering so effectively the computing infrastructure essential to our analyses. Finally, we acknowledge the enduring support for the construction and operation of the LHC and the CMS detector provided by the following funding agencies: BMBWF and FWF (Austria); FNRS and FWO (Belgium); CNPq, CAPES, FAPERJ, FAPERGS, and FAPESP (Brazil); MES (Bulgaria); CERN; CAS, MoST, and NSFC (China); COLCIENCIAS (Colombia); MSES and CSF (Croatia); RIF (Cyprus); SENESCYT (Ecuador); MoER, ERC PUT and ERDF (Estonia); Academy of Finland, MEC, and HIP (Finland); CEA and CNRS/IN2P3 (France); BMBF, DFG, and HGF (Germany); GSRT (Greece); NKFIA (Hungary); DAE and DST (India); IPM (Iran); SFI (Ireland); INFN (Italy); MSIP and NRF (Republic of Korea); MES (Latvia); LAS (Lithuania); MOE and UM (Malaysia); BUAP, CINVESTAV, CONACYT, LNS, SEP, and UASLP-FAI (Mexico); MOS (Montenegro); MBIE (New Zealand); PAEC (Pakistan); MSHE and NSC (Poland); FCT (Portugal); JINR (Dubna); MON, RosAtom, RAS, RFBR, and NRC KI (Russia); MESTD (Serbia); SEIDI, CPAN, PCTI, and FEDER (Spain); MOSTR (Sri Lanka); Swiss Funding Agencies (Switzerland); MST (Taipei); ThEPCenter, IPST, STAR, and NSTDA (Thailand); TUBITAK and TAEK (Turkey); NASU (Ukraine); STFC (United Kingdom); DOE and NSF (USA).
  
  \hyphenation{Rachada-pisek} Individuals have received support from the Marie-Curie program and the European Research Council and Horizon 2020 Grant, contract Nos.\ 675440, 724704, 752730, and 765710 (European Union); the Leventis Foundation; the A.P.\ Sloan Foundation; the Alexander von Humboldt Foundation; the Belgian Federal Science Policy Office; the Fonds pour la Formation \`a la Recherche dans l'Industrie et dans l'Agriculture (FRIA-Belgium); the Agentschap voor Innovatie door Wetenschap en Technologie (IWT-Belgium); the F.R.S.-FNRS and FWO (Belgium) under the ``Excellence of Science -- EOS" -- be.h project n.\ 30820817; the Beijing Municipal Science \& Technology Commission, No. Z191100007219010; the Ministry of Education, Youth and Sports (MEYS) of the Czech Republic; the Deutsche Forschungsgemeinschaft (DFG) under Germany's Excellence Strategy -- EXC 2121 ``Quantum Universe" -- 390833306; the Lend\"ulet (``Momentum") Program and the J\'anos Bolyai Research Scholarship of the Hungarian Academy of Sciences, the New National Excellence Program \'UNKP, the NKFIA research grants 123842, 123959, 124845, 124850, 125105, 128713, 128786, and 129058 (Hungary); the Council of Science and Industrial Research, India; the HOMING PLUS program of the Foundation for Polish Science, cofinanced from European Union, Regional Development Fund, the Mobility Plus program of the Ministry of Science and Higher Education, the National Science Center (Poland), contracts Harmonia 2014/14/M/ST2/00428, Opus 2014/13/B/ST2/02543, 2014/15/B/ST2/03998, and 2015/19/B/ST2/02861, Sonata-bis 2012/07/E/ST2/01406; the National Priorities Research Program by Qatar National Research Fund; the Ministry of Science and Higher Education, project no. 0723-2020-0041 (Russia); the Tomsk Polytechnic University Competitiveness Enhancement Program; the Programa Estatal de Fomento de la Investigaci{\'o}n Cient{\'i}fica y T{\'e}cnica de Excelencia Mar\'{\i}a de Maeztu, grant MDM-2015-0509 and the Programa Severo Ochoa del Principado de Asturias; the Thalis and Aristeia programs cofinanced by EU-ESF and the Greek NSRF; the Rachadapisek Sompot Fund for Postdoctoral Fellowship, Chulalongkorn University and the Chulalongkorn Academic into Its 2nd Century Project Advancement Project (Thailand); the Kavli Foundation; the Nvidia Corporation; the SuperMicro Corporation; the Welch Foundation, contract C-1845; and the Weston Havens Foundation (USA).
\end{acknowledgments}

\bibliography{auto_generated}

\providecommand{\href}[2]{#2}\begingroup\raggedright\begin{thebibliography}{10}%
\makeatletter
\providecommand{\hrefCMSnoop }[0]{\@secondoftwo}%
\makeatother
\providecommand{\doi}{\texttt{doi:}\begingroup \urlstyle{tt}\Url}

\bibitem{Denner:2009gj}
\hrefCMSnoop {}{A.~Denner, S.~Dittmaier, T.~Kasprzik, and A.~Muck,
  ``{Electroweak corrections to W + jet hadroproduction including leptonic
  W-boson decays}'',} \textit{ JHEP} \textbf{ 08} (2009) 075,
  \href{http://dx.doi.org/10.1088/1126-6708/2009/08/075}{\doi{10.1088/1126-6708/2009/08/075}},
\href{http://www.arXiv.org/abs/0906.1656}{\texttt{arXiv:0906.1656}}.

\bibitem{Denner:2011vu}
\hrefCMSnoop {}{A.~Denner, S.~Dittmaier, T.~Kasprzik, and A.~Muck,
  ``{Electroweak corrections to dilepton + jet production at hadron
  colliders}'',} \textit{ JHEP} \textbf{ 06} (2011) 069,
  \href{http://dx.doi.org/10.1007/JHEP06(2011)069}{\doi{10.1007/JHEP06(2011)069}},
\href{http://www.arXiv.org/abs/1103.0914}{\texttt{arXiv:1103.0914}}.

\bibitem{Denner:2012ts}
\hrefCMSnoop {}{A.~Denner, S.~Dittmaier, T.~Kasprzik, and A.~Mack,
  ``{Electroweak corrections to monojet production at the LHC}'',} \textit{
  Eur. Phys. J. C} \textbf{ 73} (2013) 2297,
  \href{http://dx.doi.org/10.1140/epjc/s10052-013-2297-x}{\doi{10.1140/epjc/s10052-013-2297-x}},
\href{http://www.arXiv.org/abs/1211.5078}{\texttt{arXiv:1211.5078}}.

\bibitem{Kallweit:2015dum}
S.~Kallweit\hrefCMSnoop {}{ {et~al.}, ``{NLO QCD+EW predictions for V + jets
  including off-shell vector-boson decays and multijet merging}'',} \textit{
  JHEP} \textbf{ 04} (2016) 021,
  \href{http://dx.doi.org/10.1007/JHEP04(2016)021}{\doi{10.1007/JHEP04(2016)021}},
  \href{http://www.arXiv.org/abs/1511.08692}{\texttt{arXiv:1511.08692}}.

\bibitem{Kuhn:2004em}
\hrefCMSnoop {}{J.~H. Kuhn, A.~Kulesza, S.~Pozzorini, and M.~Schulze,
  ``{Logarithmic electroweak corrections to hadronic Z+1 jet production at
  large transverse momentum}'',} \textit{ Phys. Lett. B} \textbf{ 609} (2005)
  277,
  \href{http://dx.doi.org/10.1016/j.physletb.2005.01.059}{\doi{10.1016/j.physletb.2005.01.059}},
\href{http://www.arXiv.org/abs/hep-ph/0408308}{\texttt{arXiv:hep-ph/0408308}}.

\bibitem{Kuhn:2005gv}
\hrefCMSnoop {}{J.~H. Kuhn, A.~Kulesza, S.~Pozzorini, and M.~Schulze,
  ``{Electroweak corrections to hadronic photon production at large transverse
  momenta}'',} \textit{ JHEP} \textbf{ 03} (2006) 059,
  \href{http://dx.doi.org/10.1088/1126-6708/2006/03/059}{\doi{10.1088/1126-6708/2006/03/059}},
\href{http://www.arXiv.org/abs/hep-ph/0508253}{\texttt{arXiv:hep-ph/0508253}}.

\bibitem{Kuhn:2005az}
\hrefCMSnoop {}{J.~H. Kuhn, A.~Kulesza, S.~Pozzorini, and M.~Schulze,
  ``{One-loop weak corrections to hadronic production of Z bosons at large
  transverse momenta}'',} \textit{ Nucl. Phys. B} \textbf{ 727} (2005) 368,
  \href{http://dx.doi.org/10.1016/j.nuclphysb.2005.08.019}{\doi{10.1016/j.nuclphysb.2005.08.019}},
\href{http://www.arXiv.org/abs/hep-ph/0507178}{\texttt{arXiv:hep-ph/0507178}}.

\bibitem{Kuhn:2007cv}
\hrefCMSnoop {}{J.~H. Kuhn, A.~Kulesza, S.~Pozzorini, and M.~Schulze,
  ``{Electroweak corrections to hadronic production of W bosons at large
  transverse momenta}'',} \textit{ Nucl. Phys. B} \textbf{ 797} (2008) 27,
  \href{http://dx.doi.org/10.1016/j.nuclphysb.2007.12.029}{\doi{10.1016/j.nuclphysb.2007.12.029}},
\href{http://www.arXiv.org/abs/0708.0476}{\texttt{arXiv:0708.0476}}.

\bibitem{Lindert:2017olm}
\hrefCMSnoop {}{J.~M. Lindert {et~al.}, ``{Precise predictions for V+jets dark
  matter backgrounds}'',} \textit{ Eur. Phys. J. C} \textbf{ 77} (2017) 829,
  \href{http://dx.doi.org/10.1140/epjc/s10052-017-5389-1}{\doi{10.1140/epjc/s10052-017-5389-1}},
\href{http://www.arXiv.org/abs/1705.04664}{\texttt{arXiv:1705.04664}}.

\bibitem{ATLAS_ZpT7TeV}
\hrefCMSnoop {}{{ATLAS Collaboration}, ``{Measurement of the transverse
  momentum distribution of Z$/\PGg^*$ bosons in proton-proton collisions at
  $\sqrt{s}=7~\TeV$ with the ATLAS detector}'',} \textit{ Phys. Lett. B}
  \textbf{ 705} (2011) 415,
  \href{http://dx.doi.org/10.1016/j.physletb.2011.10.018}{\doi{10.1016/j.physletb.2011.10.018}},
\href{http://www.arXiv.org/abs/1107.2381}{\texttt{arXiv:1107.2381}}.

\bibitem{ATLAS_ZptEta7TeV}
\hrefCMSnoop {}{{ATLAS Collaboration}, ``{Measurement of the Z$/\PGg^*$ boson
  transverse momentum distribution in $\Pp\Pp$ collisions at $\sqrt{s}=7~\TeV$
  with the ATLAS detector}'',} \textit{ JHEP} \textbf{ 09} (2014) 145,
  \href{http://dx.doi.org/10.1007/JHEP09(2014)145}{\doi{10.1007/JHEP09(2014)145}},
\href{http://www.arXiv.org/abs/1406.3660}{\texttt{arXiv:1406.3660}}.

\bibitem{Aad:2015auj}
\hrefCMSnoop {}{{ATLAS Collaboration}, ``{Measurement of the transverse
  momentum and $\phi ^*_{\eta }$ distributions of Drell-Yan lepton pairs in
  proton-proton collisions at $\sqrt{s}=8~\TeV$ with the ATLAS detector}'',}
  \textit{ Eur. Phys. J. C} \textbf{ 76} (2016) 291,
  \href{http://dx.doi.org/10.1140/epjc/s10052-016-4070-4}{\doi{10.1140/epjc/s10052-016-4070-4}},
\href{http://www.arXiv.org/abs/1512.02192}{\texttt{arXiv:1512.02192}}.

\bibitem{Aad:2019wmn}
\hrefCMSnoop {}{{ATLAS Collaboration}, ``{Measurement of the transverse
  momentum distribution of Drell\textendash{}Yan lepton pairs in
  proton\textendash{}proton collisions at $\sqrt{s}=13$ TeV with the ATLAS
  detector}'',} \textit{ Eur. Phys. J. C} \textbf{ 80} (2020) 616,
  \href{http://dx.doi.org/10.1140/epjc/s10052-020-8001-z}{\doi{10.1140/epjc/s10052-020-8001-z}},
  \href{http://www.arXiv.org/abs/1912.02844}{\texttt{arXiv:1912.02844}}.

\bibitem{CMS_ZpT7TeV}
\hrefCMSnoop {}{{CMS Collaboration}, ``{Measurement of the rapidity and
  transverse momentum distributions of Z bosons in $\Pp\Pp$ collisions at
  $\sqrt{s}=7~\TeV$}'',} \textit{ Phys. Rev. D} \textbf{ 85} (2012) 032002,
  \href{http://dx.doi.org/10.1103/PhysRevD.85.032002}{\doi{10.1103/PhysRevD.85.032002}},
  \href{http://www.arXiv.org/abs/1110.4973}{\texttt{arXiv:1110.4973}}.

\bibitem{CMS_ZpT8TeV}
\hrefCMSnoop {}{{CMS Collaboration}, ``{Measurement of the Z boson differential
  cross section in transverse momentum and rapidity in proton-proton collisions
  at 8\TeV}'',} \textit{ Phys. Lett. B} \textbf{ 749} (2015) 187,
  \href{http://dx.doi.org/10.1016/j.physletb.2015.07.065}{\doi{10.1016/j.physletb.2015.07.065}},
\href{http://www.arXiv.org/abs/1504.03511}{\texttt{arXiv:1504.03511}}.

\bibitem{CMS:2014jea}
\hrefCMSnoop {}{{CMS Collaboration}, ``{Measurements of differential and
  double-differential Drell-Yan cross sections in proton-proton collisions at
  $\sqrt{s}=8~\TeV$}'',} \textit{ Eur. Phys. J. C} \textbf{ 75} (2015) 147,
  \href{http://dx.doi.org/10.1140/epjc/s10052-015-3364-2}{\doi{10.1140/epjc/s10052-015-3364-2}},
\href{http://www.arXiv.org/abs/1412.1115}{\texttt{arXiv:1412.1115}}.

\bibitem{CMS_Z_13TeV}
\hrefCMSnoop {}{{CMS Collaboration}, ``{Measurements of differential Z boson
  production cross sections in proton-proton collisions at
  $\sqrt{s}=13~\TeV$}'',} \textit{ JHEP} \textbf{ 12} (2019) 061,
  \href{http://dx.doi.org/10.1007/JHEP12(2019)061}{\doi{10.1007/JHEP12(2019)061}},
\href{http://www.arXiv.org/abs/1909.04133}{\texttt{arXiv:1909.04133}}.

\bibitem{Aaboud:2017buf}
\hrefCMSnoop {}{{ATLAS Collaboration}, ``{Measurement of detector-corrected
  observables sensitive to the anomalous production of events with jets and
  large missing transverse momentum in $pp$ collisions at ${\sqrt{s}=13}$ TeV
  using the ATLAS detector}'',} \textit{ Eur. Phys. J. C} \textbf{ 77} (2017),
  no.~11, 765,
  \href{http://dx.doi.org/10.1140/epjc/s10052-017-5315-6}{\doi{10.1140/epjc/s10052-017-5315-6}},
  \href{http://www.arXiv.org/abs/1707.03263}{\texttt{arXiv:1707.03263}}.

\bibitem{Aaboud:2019lgy}
\hrefCMSnoop {}{{ATLAS Collaboration}, ``{Measurement of $ZZ$ production in the
  $\ell\ell\nu\nu$ final state with the ATLAS detector in $pp$ collisions at
  $\sqrt{s} = 13$ TeV}'',} \textit{ JHEP} \textbf{ 10} (2019) 127,
  \href{http://dx.doi.org/10.1007/JHEP10(2019)127}{\doi{10.1007/JHEP10(2019)127}},
  \href{http://www.arXiv.org/abs/1905.07163}{\texttt{arXiv:1905.07163}}.

\bibitem{Sirunyan:2017jix}
\hrefCMSnoop {}{{CMS Collaboration}, ``{Search for new physics in final states
  with an energetic jet or a hadronically decaying $W$ or $Z$ boson and
  transverse momentum imbalance at $\sqrt{s}=13\text{ }\text{
  }\mathrm{TeV}$}'',} \textit{ Phys. Rev. D} \textbf{ 97} (2018) 092005,
  \href{http://dx.doi.org/10.1103/PhysRevD.97.092005}{\doi{10.1103/PhysRevD.97.092005}},
  \href{http://www.arXiv.org/abs/1712.02345}{\texttt{arXiv:1712.02345}}.

\bibitem{Sirunyan:2018owy}
\hrefCMSnoop {}{{CMS Collaboration}, ``{Search for invisible decays of a Higgs
  boson produced through vector boson fusion in proton-proton collisions at
  $\sqrt{s} =$ 13 TeV}'',} \textit{ Phys. Lett. B} \textbf{ 793} (2019) 520,
  \href{http://dx.doi.org/10.1016/j.physletb.2019.04.025}{\doi{10.1016/j.physletb.2019.04.025}},
  \href{http://www.arXiv.org/abs/1809.05937}{\texttt{arXiv:1809.05937}}.

\bibitem{Sirunyan:2019ctn}
\hrefCMSnoop {}{{CMS Collaboration}, ``{Search for supersymmetry in
  proton-proton collisions at 13 TeV in final states with jets and missing
  transverse momentum}'',} \textit{ JHEP} \textbf{ 10} (2019) 244,
  \href{http://dx.doi.org/10.1007/JHEP10(2019)244}{\doi{10.1007/JHEP10(2019)244}},
  \href{http://www.arXiv.org/abs/1908.04722}{\texttt{arXiv:1908.04722}}.

\bibitem{hepdata}
\hrefCMSnoop {}{``HEPData record for this analysis''.}
  \href{http://dx.doi.org/10.17182/hepdata.96028}{\doi{10.17182/hepdata.96028}}.

\bibitem{Chatrchyan:2008aa}
\hrefCMSnoop {}{{CMS Collaboration}, ``{The CMS experiment at the CERN LHC}'',}
  \textit{ JINST} \textbf{ 3} (2008) S08004,
\href{http://dx.doi.org/10.1088/1748-0221/3/08/S08004}{\doi{10.1088/1748-0221/3/08/S08004}}.

\bibitem{Khachatryan:2016bia}
\hrefCMSnoop {}{{CMS Collaboration}, ``{The CMS trigger system}'',} \textit{
  JINST} \textbf{ 12} (2017) P01020,
  \href{http://dx.doi.org/10.1088/1748-0221/12/01/P01020}{\doi{10.1088/1748-0221/12/01/P01020}},
\href{http://www.arXiv.org/abs/1609.02366}{\texttt{arXiv:1609.02366}}.

\bibitem{Sirunyan:2017ulk}
\hrefCMSnoop {}{{CMS Collaboration}, ``{Particle-flow reconstruction and global
  event description with the CMS detector}'',} \textit{ JINST} \textbf{ 12}
  (2017) P10003,
  \href{http://dx.doi.org/10.1088/1748-0221/12/10/P10003}{\doi{10.1088/1748-0221/12/10/P10003}},
\href{http://www.arXiv.org/abs/1706.04965}{\texttt{arXiv:1706.04965}}.

\bibitem{antikt}
\hrefCMSnoop {}{{M. Cacciari and G. P. Salam and G. Soyez}, ``{The anti-\kt jet
  clustering algorithm}'',} \textit{ JHEP} \textbf{ 04} (2008) 063,
  \href{http://dx.doi.org/10.1088/1126-6708/2008/04/063}{\doi{10.1088/1126-6708/2008/04/063}},
  \href{http://www.arXiv.org/abs/0802.1189}{\texttt{arXiv:0802.1189}}.

\bibitem{Cacciari:2011ma}
\hrefCMSnoop {}{M.~Cacciari, G.~P. Salam, and G.~Soyez, ``{FastJet user
  manual}'',} \textit{ Eur. Phys. J. C} \textbf{ 72} (2012) 1896,
  \href{http://dx.doi.org/10.1140/epjc/s10052-012-1896-2}{\doi{10.1140/epjc/s10052-012-1896-2}},
\href{http://www.arXiv.org/abs/1111.6097}{\texttt{arXiv:1111.6097}}.

\bibitem{Khachatryan:2016kdb}
\hrefCMSnoop {}{{CMS Collaboration}, ``{Jet energy scale and resolution in the
  CMS experiment in pp collisions at 8 TeV}'',} \textit{ JINST} \textbf{ 12}
  (2017) P02014,
  \href{http://dx.doi.org/10.1088/1748-0221/12/02/P02014}{\doi{10.1088/1748-0221/12/02/P02014}},
  \href{http://www.arXiv.org/abs/1607.03663}{\texttt{arXiv:1607.03663}}.

\bibitem{CMS-DP-2020-019}
\href {https://cds.cern.ch/record/2715872}{{CMS Collaboration}, ``{Jet energy
  scale and resolution performance with 13 $\TeV$ data collected by CMS in
  2016-2018}'',} CMS Detector Performance Summary CMS-DP-2020-019, 2020.

\bibitem{Sirunyan:2020foa}
\hrefCMSnoop {}{{CMS Collaboration}, ``{Pileup mitigation at CMS in 13 TeV
  data}'',} \textit{ JINST} \textbf{ 15} (2020) P09018,
  \href{http://dx.doi.org/10.1088/1748-0221/15/09/P09018}{\doi{10.1088/1748-0221/15/09/P09018}},
  \href{http://www.arXiv.org/abs/2003.00503}{\texttt{arXiv:2003.00503}}.

\bibitem{Sirunyan:2019kia}
\hrefCMSnoop {}{{CMS Collaboration}, ``{Performance of missing transverse
  momentum reconstruction in proton-proton collisions at $\sqrt{s}=13~\TeV$
  using the CMS detector}'',} \textit{ JINST} \textbf{ 14} (2019) P07004,
  \href{http://dx.doi.org/10.1088/1748-0221/14/07/P07004}{\doi{10.1088/1748-0221/14/07/P07004}},
\href{http://www.arXiv.org/abs/1903.06078}{\texttt{arXiv:1903.06078}}.

\bibitem{Sirunyan:2017ezt}
\hrefCMSnoop {}{{CMS Collaboration}, ``{Identification of heavy-flavour jets
  with the CMS detector in pp collisions at 13 TeV}'',} \textit{ JINST}
  \textbf{ 13} (2018) P05011,
  \href{http://dx.doi.org/10.1088/1748-0221/13/05/P05011}{\doi{10.1088/1748-0221/13/05/P05011}},
\href{http://www.arXiv.org/abs/1712.07158}{\texttt{arXiv:1712.07158}}.

\bibitem{Khachatryan:2015hwa}
\hrefCMSnoop {}{{CMS Collaboration}, ``{Performance of electron reconstruction
  and selection with the CMS detector in proton-proton collisions at
  $\sqrt{s}=8~\TeV$}'',} \textit{ JINST} \textbf{ 10} (2015) P06005,
  \href{http://dx.doi.org/10.1088/1748-0221/10/06/P06005}{\doi{10.1088/1748-0221/10/06/P06005}},
\href{http://www.arXiv.org/abs/1502.02701}{\texttt{arXiv:1502.02701}}.

\bibitem{Sirunyan:2018fpa}
\hrefCMSnoop {}{{CMS Collaboration}, ``Performance of the {CMS} muon detector
  and muon reconstruction with proton-proton collisions at
  {$\sqrt{s}=13\TeV$}'',} \textit{ JINST} \textbf{ 13} (2018) P06015,
  \href{http://dx.doi.org/10.1088/1748-0221/13/06/P06015}{\doi{10.1088/1748-0221/13/06/P06015}},
\href{http://www.arXiv.org/abs/1804.04528}{\texttt{arXiv:1804.04528}}.

\bibitem{Cacciari:2007fd}
\hrefCMSnoop {}{M.~Cacciari and G.~P. Salam, ``{Pileup subtraction using jet
  areas}'',} \textit{ Phys. Lett. B} \textbf{ 659} (2008) 119,
  \href{http://dx.doi.org/10.1016/j.physletb.2007.09.077}{\doi{10.1016/j.physletb.2007.09.077}},
\href{http://www.arXiv.org/abs/0707.1378}{\texttt{arXiv:0707.1378}}.

\bibitem{Sirunyan:2018egh}
\hrefCMSnoop {}{{CMS Collaboration}, ``{Measurements of properties of the Higgs
  boson decaying to a W boson pair in pp collisions at $\sqrt{s}=$ 13 TeV}'',}
  \textit{ Phys. Lett. B} \textbf{ 791} (2019) 96,
  \href{http://dx.doi.org/10.1016/j.physletb.2018.12.073}{\doi{10.1016/j.physletb.2018.12.073}},
\href{http://www.arXiv.org/abs/1806.05246}{\texttt{arXiv:1806.05246}}.

\bibitem{Khachatryan:2015dfa}
\hrefCMSnoop {}{{CMS Collaboration}, ``{Reconstruction and identification of
  $\tau$ lepton decays to hadrons and $\nu_\tau$ at CMS}'',} \textit{ JINST}
  \textbf{ 11} (2016) P01019,
  \href{http://dx.doi.org/10.1088/1748-0221/11/01/P01019}{\doi{10.1088/1748-0221/11/01/P01019}},
\href{http://www.arXiv.org/abs/1510.07488}{\texttt{arXiv:1510.07488}}.

\bibitem{Sirunyan:2018pgf}
\hrefCMSnoop {}{{CMS Collaboration}, ``Performance of reconstruction and
  identification of $\tau$ leptons decaying to hadrons and $\nu_\tau$ in pp
  collisions at {$\sqrt{s}=$ 13 TeV}'',} \textit{ JINST} \textbf{ 13} (2018)
  P10005,
  \href{http://dx.doi.org/10.1088/1748-0221/13/10/P10005}{\doi{10.1088/1748-0221/13/10/P10005}},
  \href{http://www.arXiv.org/abs/1809.02816}{\texttt{arXiv:1809.02816}}.

\bibitem{CMS:EGM-14-001}
\hrefCMSnoop {}{{CMS Collaboration}, ``{Performance of photon reconstruction
  and identification with the CMS detector in proton-proton collisions at
  $\sqrt{s}=8~\TeV$}'',} \textit{ JINST} \textbf{ 10} (2015) P08010,
  \href{http://dx.doi.org/10.1088/1748-0221/10/08/P08010}{\doi{10.1088/1748-0221/10/08/P08010}},
\href{http://www.arXiv.org/abs/1502.02702}{\texttt{arXiv:1502.02702}}.

\bibitem{Alwall:2014hca}
J.~Alwall\hrefCMSnoop {}{ {et~al.}, ``{The automated computation of tree-level
  and next-to-leading order differential cross sections, and their matching to
  parton shower simulations}'',} \textit{ JHEP} \textbf{ 07} (2014) 079,
  \href{http://dx.doi.org/10.1007/JHEP07(2014)079}{\doi{10.1007/JHEP07(2014)079}},
\href{http://www.arXiv.org/abs/1405.0301}{\texttt{arXiv:1405.0301}}.

\bibitem{Kallweit:2014xda}
S.~Kallweit\hrefCMSnoop {}{ {et~al.}, ``{NLO electroweak automation and precise
  predictions for W+multijet production at the LHC}'',} \textit{ JHEP} \textbf{
  04} (2015) 012,
  \href{http://dx.doi.org/10.1007/JHEP04(2015)012}{\doi{10.1007/JHEP04(2015)012}},
  \href{http://www.arXiv.org/abs/1412.5157}{\texttt{arXiv:1412.5157}}.

\bibitem{Nason:2004rx}
\hrefCMSnoop {}{P.~Nason, ``{A new method for combining NLO QCD with shower
  Monte Carlo algorithms}'',} \textit{ JHEP} \textbf{ 11} (2004) 040,
  \href{http://dx.doi.org/10.1088/1126-6708/2004/11/040}{\doi{10.1088/1126-6708/2004/11/040}},
\href{http://www.arXiv.org/abs/hep-ph/0409146}{\texttt{arXiv:hep-ph/0409146}}.

\bibitem{Frixione:2007vw}
\hrefCMSnoop {}{S.~Frixione, P.~Nason, and C.~Oleari, ``{Matching NLO QCD
  computations with parton shower simulations: the POWHEG method}'',} \textit{
  JHEP} \textbf{ 11} (2007) 070,
  \href{http://dx.doi.org/10.1088/1126-6708/2007/11/070}{\doi{10.1088/1126-6708/2007/11/070}},
\href{http://www.arXiv.org/abs/0709.2092}{\texttt{arXiv:0709.2092}}.

\bibitem{powheg:2010}
\hrefCMSnoop {}{S.~Alioli, P.~Nason, C.~Oleari, and E.~Re, ``{A general
  framework for implementing NLO calculations in shower Monte Carlo programs:
  the POWHEG BOX}'',} \textit{ JHEP} \textbf{ 06} (2010) 043,
  \href{http://dx.doi.org/10.1007/JHEP06(2010)043}{\doi{10.1007/JHEP06(2010)043}},
  \href{http://www.arXiv.org/abs/1002.2581}{\texttt{arXiv:1002.2581}}.

\bibitem{Nason:2013ydw}
\hrefCMSnoop {}{P.~Nason and G.~Zanderighi, ``{$\PW^+\PW^-$ , $\PW\cPZ$ and
  $\cPZ\cPZ$ production in the POWHEG-BOX-V2}'',} \textit{ Eur. Phys. J. C}
  \textbf{ 74} (2014) 2702,
  \href{http://dx.doi.org/10.1140/epjc/s10052-013-2702-5}{\doi{10.1140/epjc/s10052-013-2702-5}},
\href{http://www.arXiv.org/abs/1311.1365}{\texttt{arXiv:1311.1365}}.

\bibitem{Alioli:2008gx}
\hrefCMSnoop {}{S.~Alioli, P.~Nason, C.~Oleari, and E.~Re, ``{NLO} vector-boson
  production matched with shower in {POWHEG}'',} \textit{ JHEP} \textbf{ 07}
  (2008) 060,
  \href{http://dx.doi.org/10.1088/1126-6708/2008/07/060}{\doi{10.1088/1126-6708/2008/07/060}},
\href{http://www.arXiv.org/abs/0805.4802}{\texttt{arXiv:0805.4802}}.

\bibitem{Alioli:2008tz}
\hrefCMSnoop {}{S.~Alioli, P.~Nason, C.~Oleari, and E.~Re, ``{NLO Higgs boson
  production via gluon fusion matched with shower in POWHEG}'',} \textit{ JHEP}
  \textbf{ 04} (2009) 002,
  \href{http://dx.doi.org/10.1088/1126-6708/2009/04/002}{\doi{10.1088/1126-6708/2009/04/002}},
\href{http://www.arXiv.org/abs/0812.0578}{\texttt{arXiv:0812.0578}}.

\bibitem{Sjostrand:2014zea}
T.~Sj{\"o}strand\hrefCMSnoop {}{ {et~al.}, ``{An introduction to PYTHIA
  8.2}'',} \textit{ Comput. Phys. Commun.} \textbf{ 191} (2015) 159,
  \href{http://dx.doi.org/10.1016/j.cpc.2015.01.024}{\doi{10.1016/j.cpc.2015.01.024}},
\href{http://www.arXiv.org/abs/1410.3012}{\texttt{arXiv:1410.3012}}.

\bibitem{Khachatryan:2015pea}
\hrefCMSnoop {}{{CMS Collaboration}, ``{Event generator tunes obtained from
  underlying event and multiparton scattering measurements}'',} \textit{ Eur.
  Phys. J. C} \textbf{ 76} (2016) 155,
  \href{http://dx.doi.org/10.1140/epjc/s10052-016-3988-x}{\doi{10.1140/epjc/s10052-016-3988-x}},
\href{http://www.arXiv.org/abs/1512.00815}{\texttt{arXiv:1512.00815}}.

\bibitem{Mangano:2006rw}
\hrefCMSnoop {}{M.~L. Mangano, M.~Moretti, F.~Piccinini, and M.~Treccani,
  ``{Matching matrix elements and shower evolution for top-quark production in
  hadronic collisions}'',} \textit{ JHEP} \textbf{ 01} (2007) 013,
  \href{http://dx.doi.org/10.1088/1126-6708/2007/01/013}{\doi{10.1088/1126-6708/2007/01/013}},
\href{http://www.arXiv.org/abs/hep-ph/0611129}{\texttt{arXiv:hep-ph/0611129}}.

\bibitem{Frederix:2012ps}
\hrefCMSnoop {}{R.~Frederix and S.~Frixione, ``{Merging meets matching in
  MC@NLO}'',} \textit{ JHEP} \textbf{ 12} (2012) 061,
  \href{http://dx.doi.org/10.1007/JHEP12(2012)061}{\doi{10.1007/JHEP12(2012)061}},
\href{http://www.arXiv.org/abs/1209.6215}{\texttt{arXiv:1209.6215}}.

\bibitem{Ball:2014uwa}
\hrefCMSnoop {}{{NNPDF} Collaboration, ``{Parton distributions for the LHC Run
  II}'',} \textit{ JHEP} \textbf{ 04} (2015) 040,
  \href{http://dx.doi.org/10.1007/JHEP04(2015)040}{\doi{10.1007/JHEP04(2015)040}},
\href{http://www.arXiv.org/abs/1410.8849}{\texttt{arXiv:1410.8849}}.

\bibitem{Agostinelli:2002hh}
\hrefCMSnoop {}{{\GEANTfour} Collaboration, ``{$\GEANTfour$ --- a simulation
  toolkit}'',} \textit{ Nucl. Instrum. Meth. A} \textbf{ 506} (2003) 250,
\href{http://dx.doi.org/10.1016/S0168-9002(03)01368-8}{\doi{10.1016/S0168-9002(03)01368-8}}.

\bibitem{Butterworth:2015oua}
\hrefCMSnoop {}{J.~Butterworth {et~al.}, ``{PDF4LHC recommendations for LHC Run
  II}'',} \textit{ J. Phys. G} \textbf{ 43} (2016) 023001,
  \href{http://dx.doi.org/10.1088/0954-3899/43/2/023001}{\doi{10.1088/0954-3899/43/2/023001}},
\href{http://www.arXiv.org/abs/1510.03865}{\texttt{arXiv:1510.03865}}.

\bibitem{Czakon:2017wor}
M.~Czakon\hrefCMSnoop {}{ {et~al.}, ``{Top-pair production at the LHC through
  NNLO QCD and NLO EW}'',} \textit{ JHEP} \textbf{ 10} (2017) 186,
  \href{http://dx.doi.org/10.1007/JHEP10(2017)186}{\doi{10.1007/JHEP10(2017)186}},
  \href{http://www.arXiv.org/abs/1705.04105}{\texttt{arXiv:1705.04105}}.

\bibitem{Khachatryan:2016mnb}
\hrefCMSnoop {}{{CMS Collaboration}, ``{Measurement of differential cross
  sections for top quark pair production using the lepton+jets final state in
  proton-proton collisions at 13 TeV}'',} \textit{ Phys. Rev. D} \textbf{ 95}
  (2017) 092001,
  \href{http://dx.doi.org/10.1103/PhysRevD.95.092001}{\doi{10.1103/PhysRevD.95.092001}},
  \href{http://www.arXiv.org/abs/1610.04191}{\texttt{arXiv:1610.04191}}.

\bibitem{Khachatryan:2015uqb}
\hrefCMSnoop {}{{CMS Collaboration}, ``{Measurement of the top quark pair
  production cross section in proton-proton collisions at
  $\sqrt{s}=13~\TeV$}'',} \textit{ Phys. Rev. Lett.} \textbf{ 116} (2016)
  052002,
  \href{http://dx.doi.org/10.1103/PhysRevLett.116.052002}{\doi{10.1103/PhysRevLett.116.052002}},
\href{http://www.arXiv.org/abs/1510.05302}{\texttt{arXiv:1510.05302}}.

\bibitem{Khachatryan:2016txa}
\hrefCMSnoop {}{{CMS Collaboration}, ``{Measurement of the ZZ production cross
  section and Z $\to \ell^+\ell^-\ell'^+\ell'^-$ branching fraction in pp
  collisions at $\sqrt{s}=13~\TeV$}'',} \textit{ Phys. Lett. B} \textbf{ 763}
  (2016) 280,
  \href{http://dx.doi.org/10.1016/j.physletb.2016.10.054}{\doi{10.1016/j.physletb.2016.10.054}},
\href{http://www.arXiv.org/abs/1607.08834}{\texttt{arXiv:1607.08834}}.

\bibitem{Khachatryan:2016tgp}
\hrefCMSnoop {}{{CMS Collaboration}, ``{Measurement of the WZ production cross
  section in pp collisions at $\sqrt{s}=13~\TeV$}'',} \textit{ Phys. Lett. B}
  \textbf{ 766} (2017) 268,
  \href{http://dx.doi.org/10.1016/j.physletb.2017.01.011}{\doi{10.1016/j.physletb.2017.01.011}},
\href{http://www.arXiv.org/abs/1607.06943}{\texttt{arXiv:1607.06943}}.

\bibitem{CMS-PAS-LUM-17-001}
\href {https://cds.cern.ch/record/2138682}{{CMS Collaboration}, ``{CMS}
  luminosity measurement for the 2016 data-taking period'',} CMS Physics
  Analysis Summary CMS-PAS-LUM-17-001, 2017.

\bibitem{Tikhonov:1963}
\hrefCMSnoop {}{I.~V. {Tikhonov}, ``{Solution of incorrectly formulated
  problems and the regularization method}'',} \textit{ Soviet Mathematics}
  \textbf{ 4} (1963) 1023.

\bibitem{Schmitt:2012kp}
\hrefCMSnoop {}{S.~Schmitt, ``{TUnfold: an algorithm for correcting migration
  effects in high energy physics}'',} \textit{ JINST} \textbf{ 7} (2012)
  T10003,
  \href{http://dx.doi.org/10.1088/1748-0221/7/10/T10003}{\doi{10.1088/1748-0221/7/10/T10003}},
\href{http://www.arXiv.org/abs/1205.6201}{\texttt{arXiv:1205.6201}}.

\bibitem{Ball:2017nwa}
\hrefCMSnoop {}{{NNPDF} Collaboration, ``{Parton distributions from
  high-precision collider data}'',} \textit{ Eur. Phys. J. C} \textbf{ 77}
  (2017) 663,
  \href{http://dx.doi.org/10.1140/epjc/s10052-017-5199-5}{\doi{10.1140/epjc/s10052-017-5199-5}},
\href{http://www.arXiv.org/abs/1706.00428}{\texttt{arXiv:1706.00428}}.

\bibitem{Dittmaier:2014qza}
\hrefCMSnoop {}{S.~Dittmaier, A.~Huss, and C.~Schwinn, ``{Mixed QCD-electroweak
  $\mathcal{O}(\alpha_s\alpha)$ corrections to Drell-Yan processes in the
  resonance region: pole approximation and non-factorizable corrections}'',}
  \textit{ Nucl. Phys. B} \textbf{ 885} (2014) 318,
  \href{http://dx.doi.org/10.1016/j.nuclphysb.2014.05.027}{\doi{10.1016/j.nuclphysb.2014.05.027}},
\href{http://www.arXiv.org/abs/1403.3216}{\texttt{arXiv:1403.3216}}.

\bibitem{FEWZ}
\hrefCMSnoop {}{K.~Melnikov and F.~Petriello, ``{The $W$ boson production cross
  section at the LHC through $O(\alpha^2_s)$}'',} \textit{ Phys. Rev. Lett.}
  \textbf{ 96} (2006) 231803,
  \href{http://dx.doi.org/10.1103/PhysRevLett.96.231803}{\doi{10.1103/PhysRevLett.96.231803}},
\href{http://www.arXiv.org/abs/hep-ph/0603182}{\texttt{arXiv:hep-ph/0603182}}.

\bibitem{Gavin:2010az}
\hrefCMSnoop {}{R.~Gavin, Y.~Li, F.~Petriello, and S.~Quackenbush, ``{FEWZ 2.0:
  A code for hadronic Z production at next-to-next-to-leading order}'',}
  \textit{ Comput. Phys. Commun.} \textbf{ 182} (2011) 2388,
  \href{http://dx.doi.org/10.1016/j.cpc.2011.06.008}{\doi{10.1016/j.cpc.2011.06.008}},
\href{http://www.arXiv.org/abs/1011.3540}{\texttt{arXiv:1011.3540}}.

\bibitem{Gavin:2012sy}
\hrefCMSnoop {}{R.~Gavin, Y.~Li, F.~Petriello, and S.~Quackenbush, ``{W physics
  at the LHC with FEWZ 2.1}'',} \textit{ Comput. Phys. Commun.} \textbf{ 184}
  (2013) 208,
  \href{http://dx.doi.org/10.1016/j.cpc.2012.09.005}{\doi{10.1016/j.cpc.2012.09.005}},
\href{http://www.arXiv.org/abs/1201.5896}{\texttt{arXiv:1201.5896}}.

\bibitem{Li:2012wna}
\hrefCMSnoop {}{Y.~Li and F.~Petriello, ``{Combining QCD and electroweak
  corrections to dilepton production in FEWZ}'',} \textit{ Phys. Rev. B}
  \textbf{ 86} (2012) 094034,
  \href{http://dx.doi.org/10.1103/PhysRevD.86.094034}{\doi{10.1103/PhysRevD.86.094034}},
\href{http://www.arXiv.org/abs/1208.5967}{\texttt{arXiv:1208.5967}}.

\bibitem{PDG2020}
\hrefCMSnoop {}{{Particle Data Group}, P.~A. Zyla {et~al.}, ``Review of
  particle physics'',} \textit{ Prog. Theor. Exp. Phys.} \textbf{ 2020} (2020)
  083C01,
  \href{http://dx.doi.org/10.1093/ptep/ptaa104}{\doi{10.1093/ptep/ptaa104}}.

\bibitem{Ridder:2015dxa}
A.~Gehrmann-De~Ridder\hrefCMSnoop {}{ {et~al.}, ``{Precise QCD predictions for
  the production of a Z boson in association with a hadronic jet}'',} \textit{
  Phys. Rev. Lett.} \textbf{ 117} (2016) 022001,
  \href{http://dx.doi.org/10.1103/PhysRevLett.117.022001}{\doi{10.1103/PhysRevLett.117.022001}},
\href{http://www.arXiv.org/abs/1507.02850}{\texttt{arXiv:1507.02850}}.

\end{thebibliography}\endgroup
\cleardoublepage \appendix\section{The CMS Collaboration \label{app:collab}}\begin{sloppypar}\hyphenpenalty=5000\widowpenalty=500\clubpenalty=5000\vskip\cmsinstskip
\textbf{Yerevan Physics Institute, Yerevan, Armenia}\\*[0pt]
A.M.~Sirunyan$^{\textrm{\dag}}$, A.~Tumasyan
\vskip\cmsinstskip
\textbf{Institut f\"{u}r Hochenergiephysik, Wien, Austria}\\*[0pt]
W.~Adam, T.~Bergauer, M.~Dragicevic, A.~Escalante~Del~Valle, R.~Fr\"{u}hwirth\cmsAuthorMark{1}, M.~Jeitler\cmsAuthorMark{1}, N.~Krammer, L.~Lechner, D.~Liko, I.~Mikulec, F.M.~Pitters, J.~Schieck\cmsAuthorMark{1}, R.~Sch\"{o}fbeck, M.~Spanring, S.~Templ, W.~Waltenberger, C.-E.~Wulz\cmsAuthorMark{1}, M.~Zarucki
% ...
\vskip\cmsinstskip
\dag: Deceased\\
1:  Also at Vienna University of Technology, Vienna, Austria\\
% ...
\end{sloppypar}
\end{document}